\pdfoutput=1
\documentclass[11pt]{article}

\usepackage{ACL2023}

\usepackage{times}
\usepackage{latexsym}

\usepackage[T1]{fontenc}
\usepackage[utf8]{inputenc}

\usepackage{microtype}

\usepackage{inconsolata}
\usepackage{multirow}
\usepackage{graphicx}
\usepackage{tabularx}
\usepackage{booktabs}
\usepackage[utf8]{inputenc}
\usepackage{textgreek}
\usepackage{eurosym}
\usepackage{textcomp}
\usepackage{wasysym}
\usepackage{CJKutf8}
\usepackage{seqsplit}
\usepackage{adjustbox}
\usepackage{caption}
\usepackage{subcaption}

\definecolor{mygray}{gray}{0.8}
\newcommand{\g}[1]{\fboxsep1pt\colorbox{mygray}{#1}}
\newcommand{\ul}[1]{\underline{#1}}
\newcommand{\unicode}[1]{\raisebox{-.2\height}{\includegraphics[height=1em]{unicode/#1.pdf}}}
\newcommand{\unicodebox}[1]{\setlength{\fboxsep}{0.5pt}
    \kern-0.2em
    \fcolorbox{red}{white}{\resizebox{0.3em}{0.7em}{\rotatebox{90}{\textcolor{red}{#1}}}}
    \kern-0.2em}
\newcommand{\ZWS}{\unicodebox{200b}}

\newcommand{\dataset}[0] {\texttt{BitAbuse}}

\title{\dataset{}: A Dataset of Visually Perturbed Texts for \\ Defending Phishing Attacks}

\author{
    Hanyong Lee$^{1}$, \ Chaelyn Lee$^{2}$, \ Yongjae Lee$^{3}$, Jaesung Lee$^{1,}$\thanks{\quad Corresponding author} \\
    $^{1}$ Department of Artificial Intelligence, Chung-Ang University, Seoul, South Korea\\
    $^{2}$ Korea Electronics Technology Institute, Seongnam, South Korea\\
    $^{3}$ Retrvr Inc., Seongnam, South Korea\\
    \texttt{glhy0718@gmail.com} \ \texttt{mylynchae@keti.re.kr} \ \texttt{ylee@retrvr.com} \ \texttt{curseor@cau.ac.kr}\\
}

\begin{document}
\maketitle
\begin{abstract}
Social engineering attacks, such as phishing, often target victims through visually perturbed texts to bypass security systems. The noise contained in these texts functions as an adversarial attack, designed to deceive language models and hinder their ability to accurately interpret the content. However, since it is difficult to obtain sufficient phishing cases, previous studies have used synthetic datasets that do not contain real-world cases. In this study, we propose the \dataset{} dataset, which includes real-world phishing cases, to address the limitations of previous research. Our dataset comprises a total of 325,580 visually perturbed texts. The dataset inputs are drawn from the raw corpus, consisting of visually perturbed sentences and sentences generated through an artificial perturbation process. Each input sentence is labeled with its corresponding ground truth, representing the restored, non-perturbed version. Language models trained on our proposed dataset demonstrated significantly better performance compared to previous methods, achieving an accuracy of approximately 96\%. Our analysis revealed a significant gap between real-world and synthetic examples, underscoring the value of our dataset for building reliable pre-trained models for restoration tasks. We release the \dataset{} dataset, which includes real-world phishing cases annotated with visual perturbations, to support future research in adversarial attack defense. Our code and datasets are available at \url{https://github.com/CAU-AutoML/Bitabuse}.
\end{abstract}

\section{Introduction}
{\let\thefootnote\relax\footnotetext{\color{red}WARNING: This paper contains offensive examples.}}
Social engineering attacks, including phishing, spam, pretexting, baiting, and tailgating, aim to leak confidential information by exploiting the psychological vulnerabilities of victims \citep{fi11040089}. Among them, phishing often attacks victims through texts of email, SMS, and URLs. Specifically, these phishing techniques bypass security systems such as spam filtering using visually perturbed (VP) text \citep{deng2020weaponizing, julis2020spam, boucher2022bad, unicode}, in which other characters, typically homoglyphs, replace a part of the characters in the text if the source language is English, that are nearly identical in appearance to the original characters\footnote{We will indicate such character as VP character subsequently. Similarly, VP words, VP sentences, and VP texts mean words containing VP characters, sentences containing VP words, and texts containing VP sentences, respectively.}. For example, modifying `Bitcoin' to `{\ss}itc\"{o}\u{\i}n' is an example of this technique.

Because phishing attacks based on VP texts can be prevented by restoring them to the original texts, most studies \citep{suzuki2019shamfinder, sawabe2019detection, pruthi-etal-2019-combating, imam2022ocr, keller2021bert} focused on devising an restoration method. Specifically, they modified a non-VP text dataset into a VP text dataset based on their own heuristic rules and then evaluated the performance of their restoration methods based on the synthesized dataset. These approaches are effective for identifying the weaknesses of the restoration methods, but their analysis may be biased toward their own rules because the dataset is created without regard to real-world VP texts. For example, Viper \citep{eger2019viper} always perturbs a fixed portion of characters in a sentence, which is unrealistic. Furthermore, LEGIT \citep{seth-etal-2023-learning} annotates the legibility of synthetic VP text and introduces a dataset by generating VP text that is applicable to real-world scenarios through a model that ranks transformations according to their readability. Nevertheless, research on VP text in real-world settings remains unexplored.

Although the data synthesizing strategy is helpful in circumventing the difficulty due to the lack of publicized real-world VP texts regarding phishing attacks, building a language model (LM)-based system for defending against phishing attacks only based on the synthesized dataset may be risky because there can be a gap between real-world and simulation. We argue that one way to achieve this limitation is to mix the real VP texts with the synthesized VP texts. In this case, it may be preferable that the original texts of the synthesized ones come from the same source for domain consistency. To achieve this, we propose a new dataset, namely \dataset{}, for defending phishing attacks.

Our contribution can be summarized as follows. First, based on 262,258 phishing-related emails identified from bitcoinabuse[.]com \citep{bitcoinabuse}, we created a raw corpus containing 325,580 sentences comprising 26,591 VP sentences and 298,989 non-VP English sentences. Second, based on the corpus, we created three datasets: \texttt{BitCore}, \texttt{BitViper}, and \dataset{}. Third, to depict the characteristics of our dataset, we conducted pilot studies using popular methods in this field and then compared their efficacy. We made the datasets of phishing attacks publicly available\footnote{\url{https://huggingface.co/datasets/AutoML/bitaubse}}.

\section{Related Work}

In the studies involving VP texts, obtaining sufficient data is often difficult because VP texts, usually delivered as spam emails, are not widely shared on the web. In particular, there is a lack of datasets that reflect actual phishing attack situations, and existing datasets are only valid under specific conditions or environments \citep{elsayed2018large, suzuki2019shamfinder, yazdani2020case, almuhaideb2022homoglyph}, such as internationalized domain names (IDNs.) As a result, conventional studies typically included a data synthesizing procedure with the method for restoring VP texts. Specifically, the dataset for testing the efficacy of their VP text restoration methods is synthesized by heuristic rules set in their own way.

Two notable studies regarding VP text data synthesizing are TextBugger \citep{li2019textbugger} and Viper. TextBugger is devised to generate VP texts using predefined homoglyph pairs and perturbation methods. Its goal is to degrade the performance of LMs by selecting characters in a text and replacing them with VP characters. This is useful for exposing vulnerabilities in security-sensitive tasks such as sentiment analysis \citep{pang2008opinion} or malicious content detection \citep{hou2010malicious}. Viper searches for homoglyphs and generates VP texts based on embedding techniques. This method modifies the dataset by replacing characters in the text with VP characters and induces visual disturbance based on the replacement probability.

Regarding the restoration of VP texts, conventional methods first restore malicious text using SimChar DB-based \citep{suzuki2019shamfinder}, OCR-based \citep{sawabe2019detection}, Spell Checker-based \citep{imam2022ocr}, or LM-based methods \citep{keller2021bert} and then detect malicious texts. The SimChar DB-based method automatically collects homoglyphs from the Unicode character set to detect VP characters in IDNs and restores them using a predefined restoration table. OCR-based methods were investigated to detect phishing attacks that deceive users by putting VP characters in IDNs. This method recognizes VP characters as images and converts them into the original characters. The Spell Checker-based method aims to detect images containing malicious text distributed on social networks by considering deformed characters in the text as typos and restoring them using a spell checker. The restoration strategy that combines two LMs, BERT \citep{devlin2019bert} and GPT \citep{radford2018improving}, was also considered \citep{keller2021bert}.

A common drawback of conventional studies is that the datasets used for evaluating the restoration performance of phishing attacks contain no real VP texts. As a result, the restoration performance in real-world situations may be over/underestimated, and unstable pre-trained LM models can be obtained. In this study, we create a new dataset that can contribute to phishing attack studies by collecting VP texts used in bitcoinabuse[.]com.

\begin{table*}[!t]
\centering
{
\newcolumntype{L}{>{\raggedright\arraybackslash}X}
\begin{tabularx}{\textwidth}{LL}
\toprule
\multicolumn{1}{c}{Original VP sentence} & \multicolumn{1}{c}{Ground truth (VP characters are restored)} \\
\midrule
i am going to s\g{\unicode{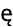}}nd out your vid\g{\unicode{0119}}o record\g{\unicode{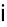}}ng to ev\g{\unicode{0119}}ry bit of yo\g{\unicode{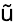}}r contacts and you c\g{\unicode{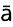}}n easily im\g{\unicode{0101}}g\g{\unicode{0456}}ne c\g{\unicode{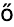}}ncerning the disgrac\g{\unicode{0119}} you will s\g{\unicode{0119}}e. & i am going to send out your video recording to every bit of your contacts and you can easily imagine concerning the disgrace you will see.\\
\midrule
\g{\unicode{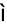}}f you w\g{\unicode{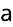}}nt to prevent th\g{\unicode{00ec}}s, tr\g{\unicode{0430}}nsfer 0.019 btc to my b\g{\unicode{00ec}}tco\g{\unicode{00ec}}n w\g{\unicode{0430}}llet (\g{\unicode{00ec}}n c\g{\unicode{0430}}se you do not know how to do \g{\unicode{00ec}}t, then wr\g{\unicode{00ec}}te to google: "buy \g{\unicode{0430}} b\g{\unicode{00ec}}tco\g{\unicode{00ec}}n"). & if you want to prevent this, transfer 0.019 btc to my bitcoin wallet (in case you do not know how to do it, then write to google: "buy a bitcoin").\\
\bottomrule
\end{tabularx}
}
\caption{\label{tab:dataset_example} Examples of VP sentences (Left) and manually restored sentences (Right) in the raw corpus (note that vowels such as 'a', 'e', 'i', 'o', and 'u' are mainly used as VP, which are highlighted in gray).}
\end{table*}

\section{\dataset{}}

We collect VP texts used in phishing attacks from the bitcoinabuse[.]com \citep{bitcoinabuse} website. The website bitcoinabuse[.]com is a platform where worldwide users can share content related to Bitcoin fraud, such as emails. The site provides data collected through user participation, making it easy to find phishing email bodies containing VP texts. Additionally, because users directly upload emails after masking personal information, it can be ensured that the data can be collected safely without privacy concerns.

We used 262,258 phishing-related emails collected from bitcoinabuse[.]com between May 16, 2017, and January 15, 2022. Detailed statistics of the raw dataset are discussed in Appendix~\ref{sec:raw_dataset_statistics}.
% Table~\ref{tab:raw_dataset} shows brief statistics of the collected email texts. We identified 262,258 phishing-related emails from bitcoinabuse[.]com between May 16, 2017, and January 15, 2022, and extracted the text bodies of these emails. The length of the email bodies averages about 417 characters, ranging from a minimum of 10 characters to a maximum of 2,000 because the platform limits the maximum number of characters to 2,000. The content of phishing-related emails was uploaded from approximately 224 countries, and the country of upload and the language of the collected text may differ.

Although our primary goal is to create a VP text dataset related to English texts, the emails collected from English-speaking countries also included non-English texts, so removing irrelevant emails was followed. However, using existing language detection models to classify emails written in English is challenging due to the presence of English VP text, and fully manual filtering is also impractical. Therefore, we utilized the BERT model \citep{devlin2019bert} with a fully connected classification layer trained to automatically classify English text. The BERT model used in this classifier has a hidden state size of 768, 12 hidden layers, and 12 attention heads. Among the 262,258 email texts, 16,598 were randomly chosen and manually labeled as English (10,024 texts) or non-English (6,574 texts), and these labeled email texts were used to train the classification model. The labeled dataset was exclusively divided into train, validation, and test sets with 13,444, 1,494, and 1,660 texts, respectively. We provide the detailed hardware specification and hyperparameter settings in Appendix~\ref{sec:filteringBERT}.

The classifier achieved an accuracy of approximately 99.28\% on 1,660 uninvolved email texts in the training phase. The trained classifier removed 84,204 non-English email texts from the 262,258 ones, resulting in 178,054 email texts for further processing. Although this process significantly accelerates the preprocessing, non-English emails may still remain because the classification is imperfect. Such non-English sentences from those email texts are removed manually during a subsequent process that will be explained later.

After rough filtering 178,054 non-English email texts, we obtained 326,732 sentences by splitting the original texts with a maximum length of 512. Because those sentences include unnecessary components, such as random character sequences, we used a series of regular expressions to remove them efficiently. The list of regular expressions we used for further preprocessing and downloadable URL links are presented in Appendix~{\ref{sec:regex_examples}. We found that the raw dataset contains a wide range of VP  characters not addressed in previous studies, such as control characters from U+0001 to U+0005, that will remain in \texttt{BitCore} and \dataset{} datasets.

% Table~\ref{tab:replace_example} shows example sentences after the preprocessing based on the regular expressions. In three examples of the table, emojis and special characters in the sentence are removed, and unusual characters are replaced with ASCII characters with the same meaning. For example, in the first example in the table, ``Left Black Lenticular Bracket (U+3010)'' and ``Right Black Lenticular Bracket (U+3011)'' were replaced with regular parentheses (U+0028, U+0029). In the second example, unprintable Unicode characters that are presented as a hex value in the red box are removed. 

To validate the restoration performance, we manually annotated the label for each character in the 326,732 sentences. 
Since manually annotating VP text is highly labor-intensive and inefficient, we extracted VP words from the VP text and manually created non-VP word labels for each VP word. These labels were then applied to the VP text to generate non-VP text labels. In cases where it was difficult to determine the label by looking only at the VP word, we referred to the original text to accurately annotate the corresponding non-VP word.
While annotating, 1,152 irrelevant sentences, such as repetitive identical characters, random character sequences, or non-English sentences missed from the previous classification process, were removed. 
% As a result, 325,580 sentences comprising 26,591 VP and 298,989 non-VP English sentences remained, forming our raw corpus.
% Tables~\ref{tab:dataset_example} and~\ref{tab:phishirest_dataset} show the examples of VP text with corresponding ground truth sentences and brief statistics of the raw corpus, respectively.
Table~\ref{tab:dataset_example} shows examples of VP text with the corresponding ground truth sentences, and brief statistics of the raw corpus are presented in Appendix~\ref{sec:raw_dataset_statistics}. Also, We created \texttt{BitCore}, \texttt{BitViper}, and \texttt{BitAbuse} datasets based on the raw corpus. Brief statistics of the three datasets are presented in Table~\ref{tab:datasets_stat} of Appendix \ref{sec:bitabuse_statistics}.
% Specifically, \texttt{BitCore} was created by simply selecting 26,591 VP sentences from the raw corpus. Next, \texttt{BitViper} was created by applying the character perturbation procedure of Viper that uses the ICEs method with a probability of 0.2 to 298,989 non-VP sentences of the raw corpus, following the same settings used in the original study for the restoration task\footnote{TextBugger is not considered here because it attacks by altering keywords in sentences for semantic classification. Thus, applying TextBugger to non-VP texts of the raw corpus requires additional work, such as labeling whether a sentence is spam, which is out of the scope of this study.}. Lastly, \dataset{} was created by merging \texttt{BitCore} and \texttt{BitViper}, resulting in the largest dataset of our study that contains both real-world and synthetic VP sentences.
% All these three datasets can be downloaded from a public repository.
% \footnote{\url{https://tobeassigned.com}, \url{https://tobeassigned.com}, \url{https://tobeassigned.com}}.

\begin{figure*}[!t]
    \centering
    \begin{subfigure}[b]{0.96\columnwidth}
        \centering
        \includegraphics[width=\textwidth]{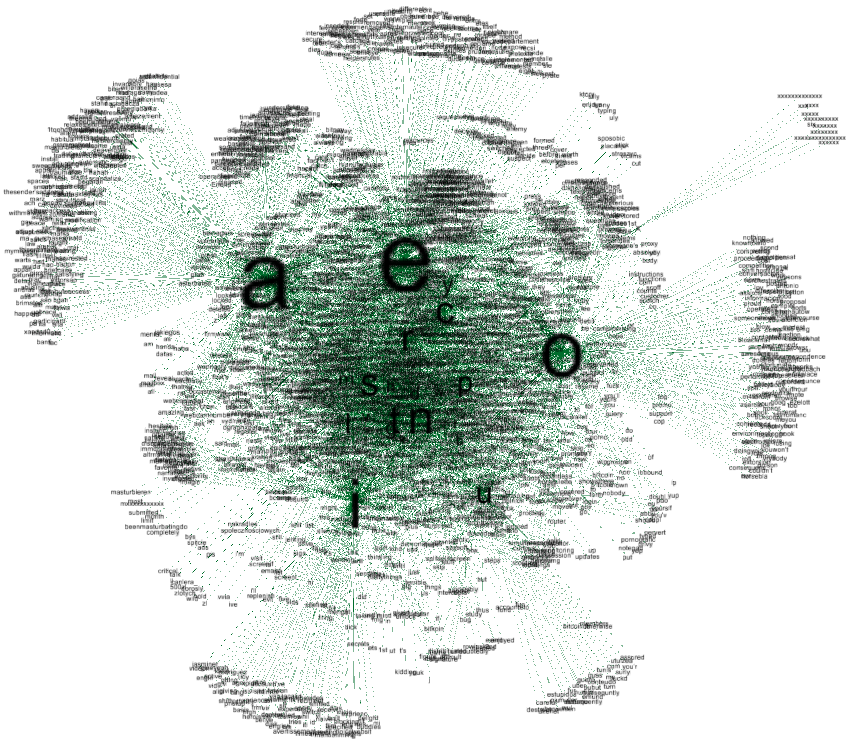}
        \caption{Complete graph of VP character-word association}
    \end{subfigure}
    \hfill
    \begin{subfigure}[b]{0.96\columnwidth}
        \centering
        \begin{subfigure}[b]{0.49\textwidth}
            \centering
            \includegraphics[width=\textwidth]{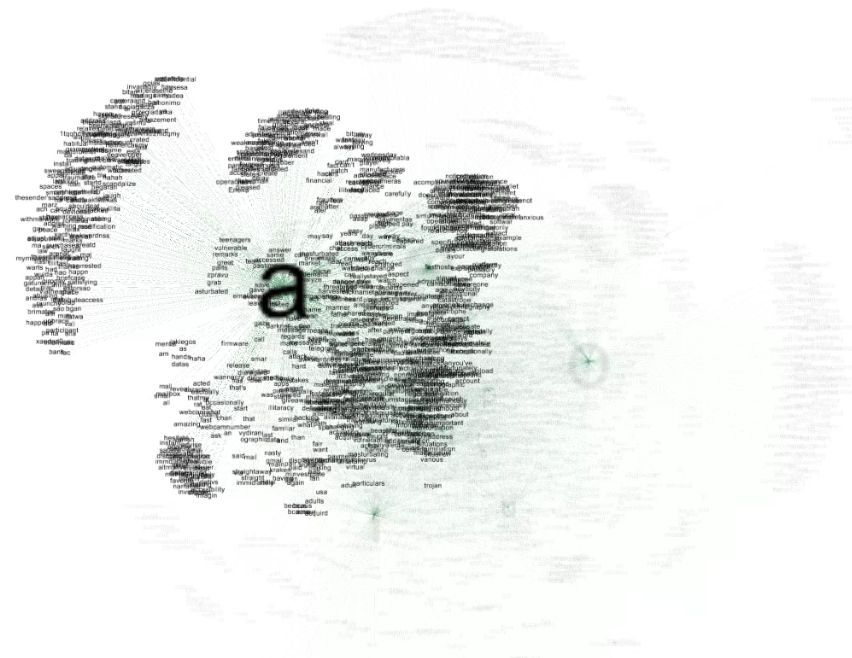}
            \caption{Subgraph regarding `a'}
        \end{subfigure}
        \hfill
        \begin{subfigure}[b]{0.49\textwidth}
            \centering
            \includegraphics[width=\textwidth]{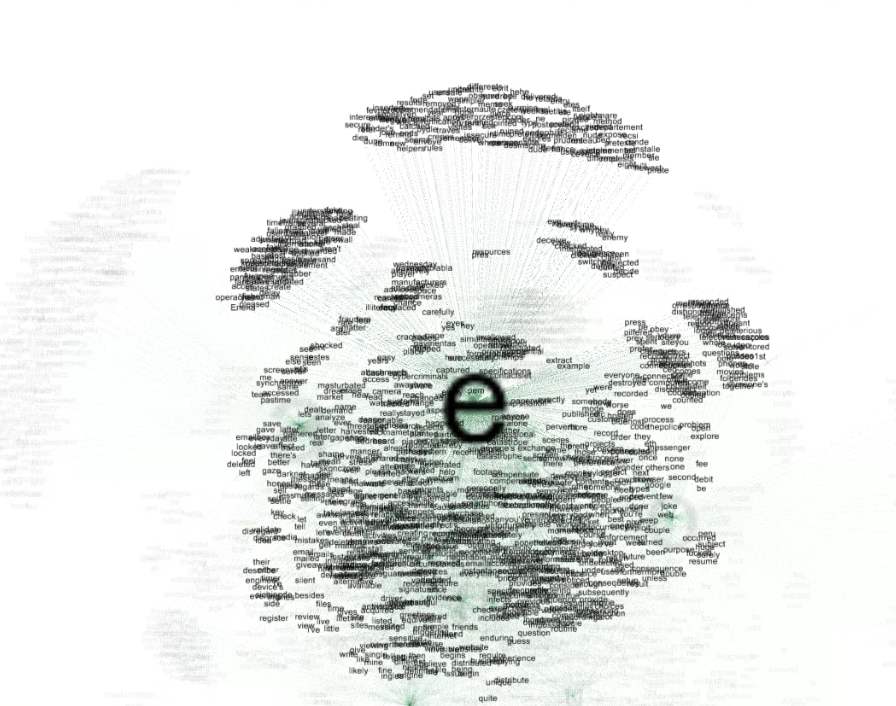}
            \caption{Subgraph regarding `e'}
        \end{subfigure}
        \vskip\baselineskip
        \begin{subfigure}[b]{0.49\textwidth}
            \centering
            \includegraphics[width=\textwidth]{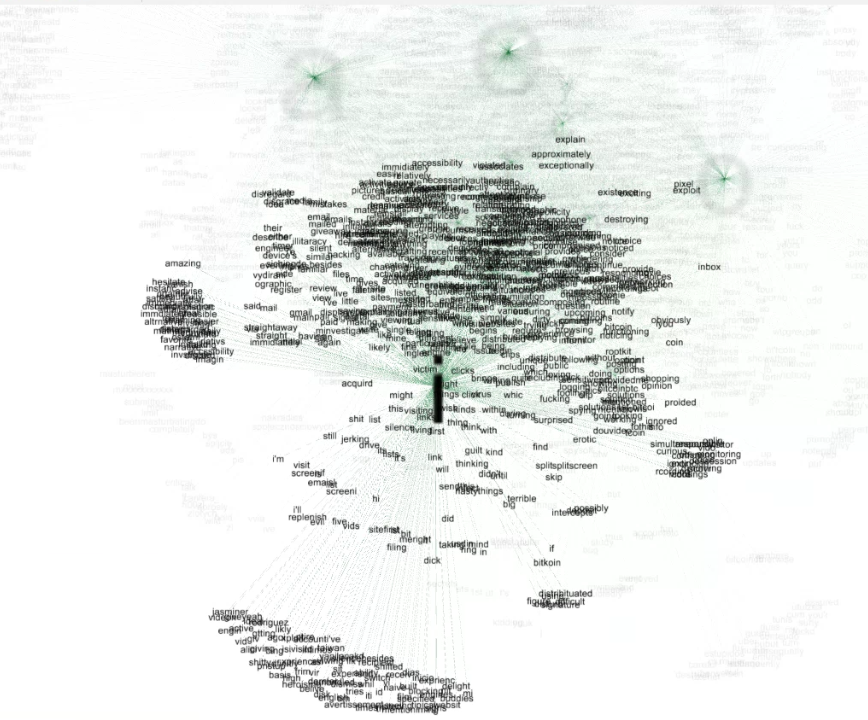}
            \caption{Subgraph regarding `i'}
        \end{subfigure}
        \hfill
        \begin{subfigure}[b]{0.49\textwidth}
            \centering
            \includegraphics[width=\textwidth]{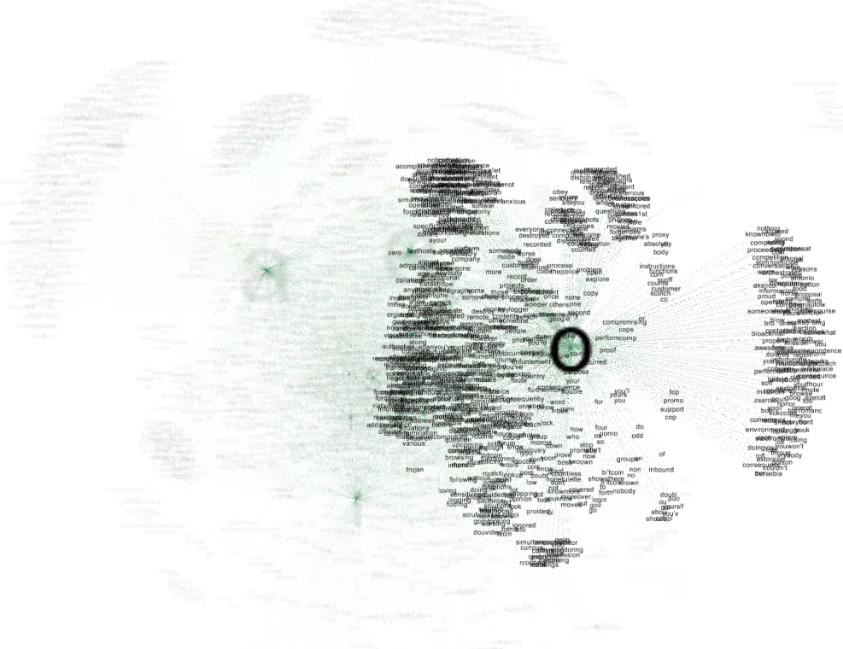}
            \caption{Subgraph regarding `o'}
        \end{subfigure}        
    \end{subfigure}
    \caption{Visualization of clustering based on VP character-word association of \texttt{BitCore} dataset. (a) overview of the obtained graph, (b) subgraph regarding VP characters of `a', (c) subgraph regarding VP characters of `e', (d) subgraph regarding VP characters of `i', and (e) subgraph regarding VP characters of `o'}
    \label{fig:bitcore_clustering}
\end{figure*}

\begin{table*}[!t]
\centering
{
\setlength{\tabcolsep}{5pt}
\small
\begin{tabularx}{\textwidth}{c l ccccc}
\toprule
\multirow{2}[3]{*}{Measure} & \multicolumn{1}{c}{\multirow{2}[3]{*}{Dataset}} & \multicolumn{5}{c}{Restoration Method} \\
\cmidrule{3-7}
& & SimChar DB & OCR & Spell Checker & Character BERT & GPT-4o mini\\
\midrule
Word & \texttt{BitCore} & $0.5515 \pm 0.0036$ & $0.6531 \pm 0.0036$ & $\underline{0.8909} \pm 0.0016$ & $\textbf{0.9984} \pm 0.0004$ & $0.7168 \pm 0.0040$ \\
Level & \texttt{BitViper} & $0.3373 \pm 0.0006$ & $0.3177 \pm 0.0006$ & $\underline{0.7133} \pm 0.0011$ & $\textbf{0.9534} \pm 0.0008$ & $0.5034 \pm 0.0009$ \\
Accuracy & \dataset{} & $0.3547 \pm 0.0010$ & $0.3446 \pm 0.0008$ & $\underline{0.7275} \pm 0.0012$ & $\textbf{0.9568} \pm 0.0006$ & $0.5196 \pm 0.0010$ \\
\midrule
Word & \texttt{BitCore} & $0.6581 \pm 0.0026$ & $0.7255 \pm 0.0022$ & $0.8734 \pm 0.0016$ & $\textbf{0.9992} \pm 0.0001$ & $\underline{0.8966} \pm 0.0023$ \\
Level & \texttt{BitViper} & $0.4708 \pm 0.0005$ & $0.4617 \pm 0.0005$ & $0.6942 \pm 0.0010$ & $\textbf{0.9294} \pm 0.0010$ & $\underline{0.7963} \pm 0.0007$ \\
Jaccard & \dataset{} & $0.4860 \pm 0.0007$ & $0.4830 \pm 0.0007$ & $0.7083 \pm 0.0009$ & $\textbf{0.9347} \pm 0.0008$ & $\underline{0.8037} \pm 0.0005$ \\
\midrule
 & \texttt{BitCore} & $0.8199 \pm 0.0011$ & $0.8860 \pm 0.0011$ & $\underline{0.9476} \pm 0.0008$ & $\textbf{0.9997} \pm 0.0000$ & $0.9328 \pm 0.0025$ \\
BLEU & \texttt{BitViper} & $0.7808 \pm 0.0003$ & $0.7748 \pm 0.0002$ & $0.8753 \pm 0.0005$ & $\textbf{0.9765} \pm 0.0004$ & $\underline{0.8919} \pm 0.0004$ \\
 & \dataset{} & $0.7838 \pm 0.0004$ & $0.7836 \pm 0.0004$ & $0.8809 \pm 0.0005$ & $\textbf{0.9782} \pm 0.0003$ & $\underline{0.8947} \pm 0.0004$ \\
\bottomrule
\end{tabularx}
}
\caption{\label{tab:comp_three}Comparison results of the five restoration methods in terms of three evaluation measures. Bold text indicates the best performance, and underlining indicates the second-best performance.}
\end{table*}

\section {Experimental Settings}
\label{sec:restorSettings}

We tested the restoration performance using Simchar DB \citep{suzuki2019shamfinder}, OCR \citep{sawabe2019detection}, Spell Checker \citep{imam2022ocr}, Character BERT-based\citep{el2020characterbert}, and GPT-4o mini-based methods \citep{openai2023gpt4} in the viewpoint of three well-known evaluation measures, such as Word Level Accuracy \citep{imam2022ocr}, Word Level Jaccard, and BLEU \citep{zeng-etal-2021-openattack}. In addition, detailed information regarding the experiments, such as the model's hyperparameters, is described in Appendix~\ref{sec:exp_details}.

\subsection{Methods}

\begin{table*}[!t]
    \centering
    \newcolumntype{L}{>{\raggedright\arraybackslash}X}
    \definecolor{mygray}{gray}{0.8}
    \small
    \begin{tabularx}{\textwidth}{c LL}
        \toprule
        Method & \multicolumn{1}{c}{Example 1} & \multicolumn{1}{c}{Example 2} \\
        \midrule
        \begin{tabular}{@{}c@{}} Original \\ Text \end{tabular} & \begin{tabular}{@{}l@{}} afte\g{\unicode{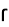}} t\g{\unicode{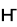}}\g{\unicode{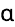}}t, i h\g{\unicode{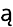}}v\g{\unicode{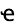}} st\g{\unicode{0105}}\g{\unicode{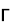}}t\g{\unicode{04bd}}d tr\g{\unicode{03b1}}c\g{\unicode{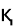}}ing \\ \g{\unicode{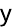}}\g{\unicode{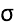}}\g{\unicode{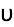}}\g{\unicode{0433}} \g{\unicode{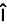}}nt\g{\unicode{04bd}}\g{\unicode{0433}}\g{\unicode{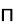}}\g{\unicode{04bd}}t \g{\unicode{03b1}}ct\g{\unicode{00ee}}v\g{\unicode{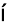}}t\g{\unicode{00ed}}\g{\unicode{04bd}}s. \end{tabular} & \begin{tabular}{@{}l@{}} \g{\unicode{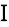}} \g{\unicode{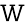}}\g{\unicode{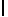}}\g{\unicode{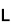}}\g{\unicode{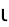}} \g{\unicode{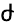}}\g{\unicode{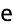}}\g{\unicode{029f}}\g{\unicode{0435}}\g{\unicode{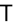}}\g{\unicode{0435}} \g{\unicode{0435}}\g{\unicode{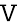}}\g{\unicode{0435}}\g{\unicode{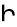}}\g{\unicode{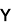}}\g{\unicode{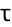}}\g{\unicode{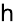}}\g{\unicode{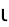}}\g{\unicode{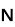}}\g{\unicode{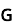}} \g{\unicode{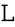}}\textquotesingle \g{\unicode{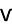}}\g{\unicode{0435}} \\ \g{\unicode{0262}}\g{\unicode{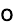}}\g{\unicode{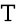}} \g{\unicode{0430}}\g{\unicode{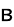}}0\g{\unicode{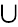}}\g{\unicode{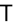}} \g{\unicode{028f}}\g{\unicode{03bf}}\g{\unicode{22c3}}. \end{tabular} \\
        \midrule
        \begin{tabular}{@{}c@{}}SimChar \\ DB \end{tabular} 
        & \begin{tabular}{@{}l@{}} after t\ul{\g{\unicode{04a5}}}\ul{\g{\unicode{03b1}}}t, i hav\ul{\g{\unicode{04bd}}} start\ul{\g{\unicode{04bd}}}d tr\ul{\g{\unicode{03b1}}}cking \\ \ul{\g{\unicode{0443}}}\ul{\g{\unicode{03c3}}}ur int\ul{\g{\unicode{04bd}}}rn\ul{\g{\unicode{04bd}}}t \ul{\g{\unicode{03b1}}}ctiviti\ul{\g{\unicode{04bd}}}s. \end{tabular} 
        & \begin{tabular}{@{}l@{}} \ul{\g{\unicode{0399}}} \ul{\g{\unicode{13b3}}}\ul{\g{\unicode{16c1}}}\ul{\g{\unicode{029f}}}\ul{\g{\unicode{03b9}}} \ul{\g{\unicode{056a}}}e\ul{\g{\unicode{029f}}}e\ul{\g{\unicode{03a4}}}e e\ul{\g{\unicode{13d9}}}e\ul{\g{\unicode{053b}}}\ul{\g{\unicode{028f}}}\ul{\g{\unicode{03c4}}}\ul{\g{\unicode{04bb}}}\ul{\g{\unicode{0269}}}\ul{\g{\unicode{0274}}}\ul{o} \ul{\g{\unicode{13de}}}\textquotesingle \ul{\g{\unicode{0475}}}e \\ \ul{o}o\ul{\g{\unicode{13a2}}} a\ul{e}\ul{0}\ul{\g{\unicode{22c3}}}\ul{\g{\unicode{0422}}} \ul{\g{\unicode{028f}}}o\ul{\g{\unicode{22c3}}}. \end{tabular} \\
        \midrule
        OCR 
        & \begin{tabular}{@{}l@{}} after that, i h\ul{q}v\ul{a} st\ul{q}rt\ul{a}d tracking your \\ \ul{t}nt\ul{a}rn\ul{a}t act\ul{t}v\ul{\g{\unicode{00ed}}}t\ul{\g{\unicode{00ed}}}\ul{a}s. \end{tabular}
        & i w\ul{\g{\unicode{16c1}}}ll \ul{\g{\unicode{056a}}}elete eve\ul{h}yth\ul{l}ng \ul{l}\textquotesingle ve got ab\ul{0}ut you. \\
        \midrule
        \begin{tabular}{@{}c@{}} Spell \\ Checker \end{tabular} 
        & \begin{tabular}{@{}l@{}} after that, i have started tracking your \\ \ul{<None>} \ul{<None>}. \end{tabular}
        & i will delete everything \ul{l}\ul{o}ve got about you.\\
        \midrule
        \begin{tabular}{@{}c@{}} Character \\ BERT \end{tabular} 
        & \begin{tabular}{@{}l@{}} after\ that,\ i\ have\ started\ tracking\ your\ \\ internet\ activities. \end{tabular}
        & i\ will\ delete\ everything\ i\textquotesingle ve\ got\ about\ you.\\
        \midrule
         \begin{tabular}{@{}c@{}} GPT-4o \\ mini \end{tabular} 
         & \begin{tabular}{@{}L@{}} after that, i have started \ul{taking} about internet activities.\end{tabular}
         & i will \ul{let you know} about \ul{what} you. \\
        \bottomrule
    \end{tabularx}
    \caption{Restoration examples of the five methods where ``<None>'' indicates that the method failed to restore the word (VP and incorrectly restored characters are highlighted in gray color and underlined, respectively.)}
    \label{tab:result_restore}
\end{table*}

% Additionally, examples of VP variations for the words with the most variations in Table~\ref{tab:extended_interest_words} are also shown in Table~\ref{tab:vpw_example} in the Appendix \ref{sec:extended_lists}.
We tested the restoration performance using five different methods. The SimChar DB-based method checks if there is an alphabetic homoglyph for each character in the Simchar Database and uses it to restore the homoglyph.
The OCR-based method was implemented by applying OCR to each character and selecting the character with the highest probability.
Spell Checker-based method entailed the segmentation of sentences into individual word units through a rule-based approach, followed by the restoration of each word using a spell checker based on Levenshtein Distance, as documented in the corresponding references \citep{norvig2016how, lison2016opensubtitles2016}. 

\paragraph{Character BERT} The Character BERT-based method employs a BERT model that processes token sequences at the character level to restore VP characters, inferring them as the original characters through the context of individual characters. In this approach, instead of relying on a standard subword tokenizer-based BERT—which is less effective when tokens contain perturbed characters—a character-level sequence approach is adopted for both input and output. This method is particularly important because attackers often modify characters within tokens to deceive victims, leading to widespread perturbations across most tokens. Standard BERT's Masked Language Model (MLM) mechanism, which relies on contextual information from surrounding tokens, struggles in such cases because the context tokens themselves may also be perturbed. Experiments with the Character BERT-based model involve training to restore and output the input VP sentence from three datasets into the corresponding restored sentence. In this process, both the input and output are sequences of character-level tokens. The training process of Character BERT was configured with a learning rate of $5 \times 10^{-5}$, a batch size of 32, and ten training epochs. Additionally, the AdamW optimizer was used with settings of $\beta_{1} = 0.9$, $\beta_{2} = 0.999$, and a $\textrm{weight\_decay} = 0$, along with a linear learning rate scheduler. The experiment shown in Table~\ref{tab:comp_crossval} uses the same hyperparameters as the previously mentioned experiment, except the number of training epochs is set to 20.

We also employed the GPT-4o mini model to assess the performance of the latest large language model on the \dataset{}. GPT-4o mini is a closed-source generative language model, and the experiment was conducted via OpenAI's inference API. To leverage the model, we designed a prompt, as detailed in Table~\ref{tab:gpt4_prompt} of Appendix~\ref{sec:exp_details}.

\subsection{Evaluations}

We evaluated each method using the three measures that were used in previous VP text restoration studies: Word Level Accuracy, Word Level Jaccard, and BLEU. Word Level Accuracy is a measure that evaluates whether the restored word matches at each word position. When $N_{c}$ represents the number of correctly restored words and $N$ represents the total number of words in each sentence, Word Level Accuracy is calculated as

$$ \textrm{Word Level Accuracy} = \frac{N_{c}}{N} .$$

The Word Level Jaccard score is calculated by forming the word set $W_{p}$ from the predicted sentence and the word set $W_{l}$ from the labeled sentence and then computing the ratio of the size of their intersection to the size of their union. Specifically, the Word Level Jaccard score is defined as

$$ \textrm{Word Level Jaccard} = \frac{|W_{p} \cap W_{l}|}{|W_{p} \cup W_{l}|} .$$

The BLEU score is calculated by constructing the character sequences $C_{p}$ of the predicted sentence and the character sequences $C_{l}$ of the labeled sentence and then calculating the precision of the $n$-grams of the two sequences by

$$ \textrm{BLEU} = \textrm{B} \times \exp\left(\sum_{n=1}^{N} w_n \log p_n\right) .$$
where $N$ and $w_n$ are the maximum length and the weight of the $n$-grams, respectively. $p_n$ represents the precision of the $n$-grams in $C_{l}$ and $C_{p}$, and $B$ is the brevity penalty used in the BLEU score calculation. In this paper, $N=4$ is used to calculate the BLEU score, and $w_n=1/N$ is set. The brevity penalty follows the standard BLEU score calculation method.

Without specific mentions, among sentences of \dataset{} dataset, 60\%, 20\%, and 20\% of them were used for training, validation, and testing, respectively. The performance of each method was evaluated on the test set, and the average performance was measured by repeating the experiments with ten random training and test set splits.

\section{Experimental Results}
\label{sec:ext_res}

We conducted exploratory data analysis on VP words, VP characters, ratios, and so on that may help devise an effective methodology for defending phishing attacks. 

\begin{figure*}[!t]
\centering
\centering
\begin{subfigure}[b]{0.32\textwidth}
    \centering
    \includegraphics[width=\textwidth]{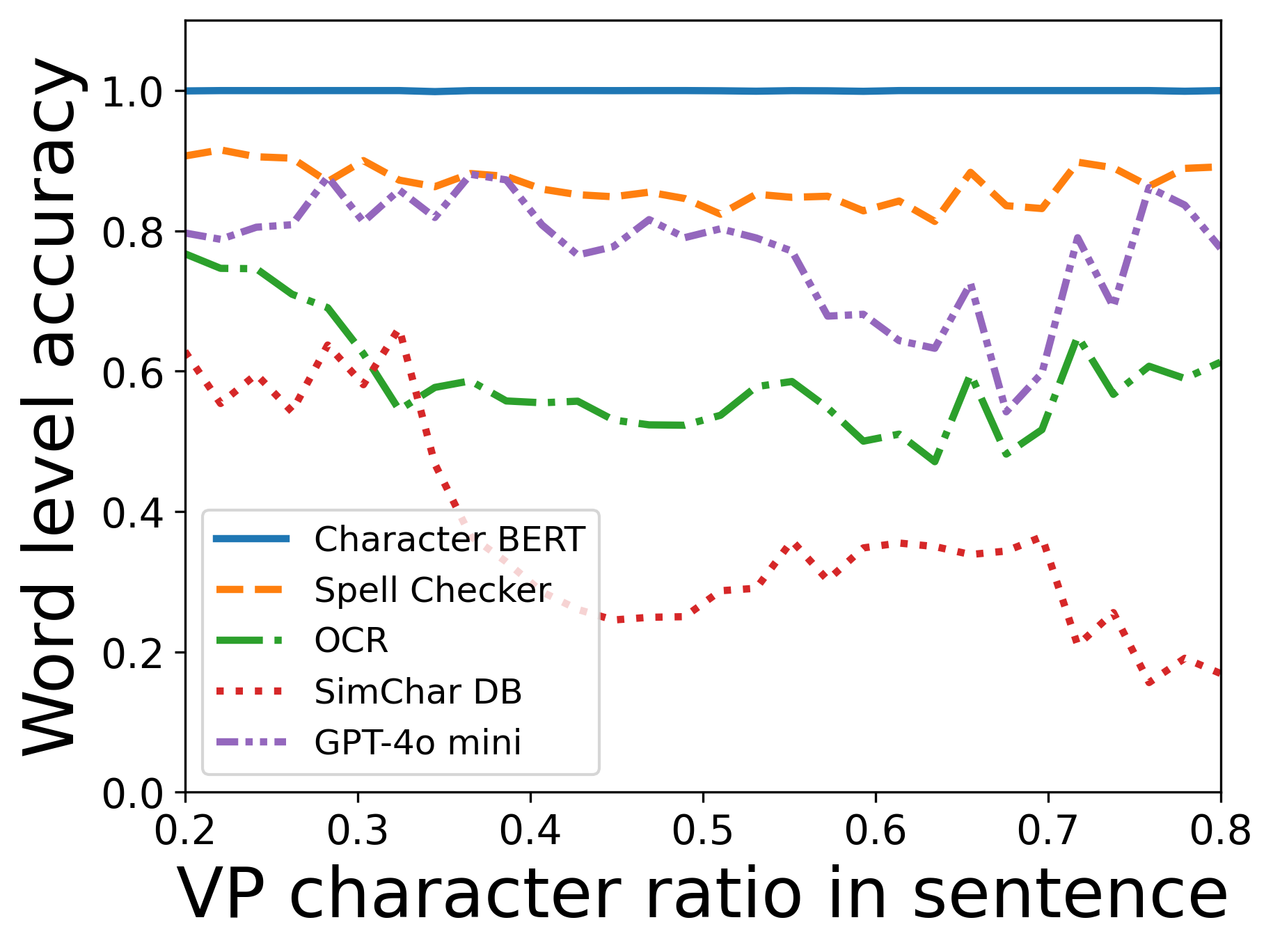}
    \caption{\texttt{BitCore} dataset}
    \label{fig:bitcore_ratio_wra}
\end{subfigure}
\hfill
\begin{subfigure}[b]{0.32\textwidth}
    \centering
    \includegraphics[width=\textwidth]{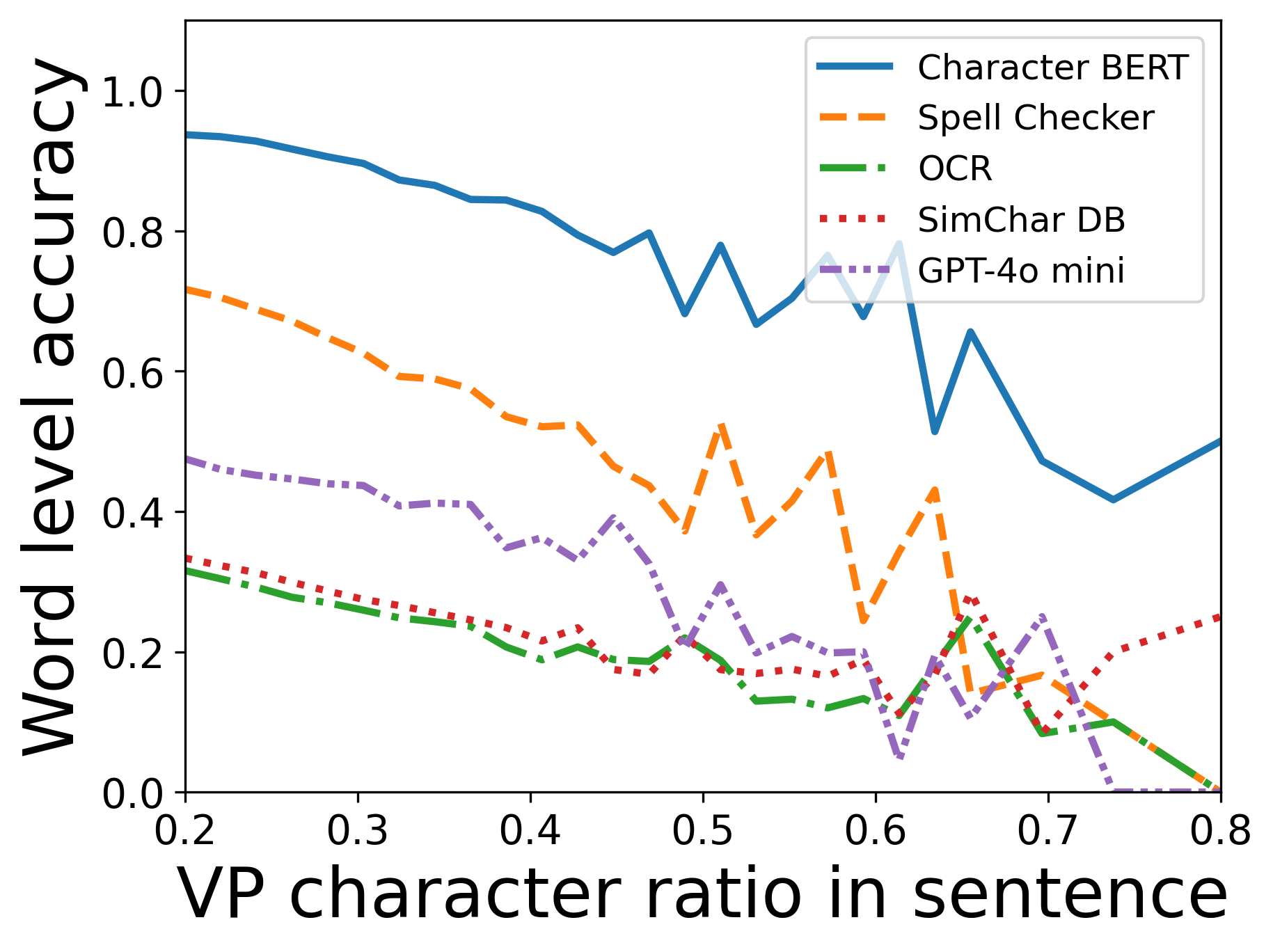}
    \caption{\texttt{BitViper} dataset}
    \label{fig:bitviper_ratio_wra}
\end{subfigure}
\hfill
\begin{subfigure}[b]{0.32\textwidth}
    \centering
    \includegraphics[width=\textwidth]{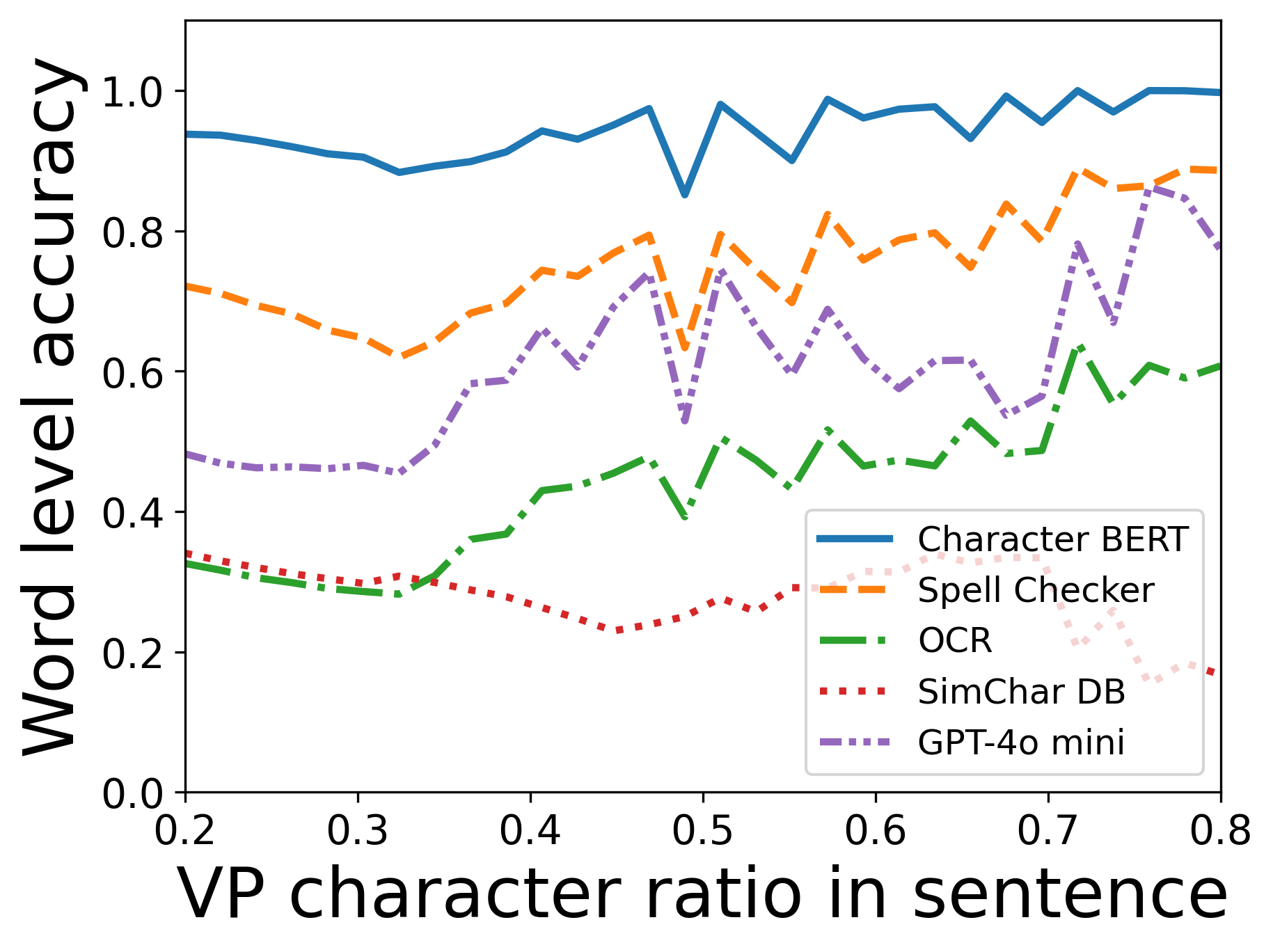}
    \caption{\dataset{} dataset}
    \label{fig:bitabuse_ratio_wra}
\end{subfigure}
\caption{Word Level Accuracy performance of each method regarding VP character ratio in each sentence}
\label{fig:homoglyph_ratio_acc}
\end{figure*}

\begin{table*}[!t]
\centering
\begin{tabularx}{\textwidth}{c l cccc}
\toprule
\multirow{2}{*}{Measure} & \multicolumn{1}{c}{\multirow{2}{*}{Dataset}} & \multicolumn{4}{c}{Training / Validation / Test ratio (\%)} \\
\cline{3-6}
& & 1 / 20 / 79 & 5 / 20 / 75 & 10 / 20 / 70 & 20 / 20 / 60 \\
\midrule
Word & \texttt{BitCore} & $0.9704 \pm 0.0047$ & $0.9917 \pm 0.0016$ & $0.9957 \pm 0.0003$ & $\textbf{0.9975} \pm 0.0007$ \\
Level & \texttt{BitViper} & $0.5318 \pm 0.0013$ & $0.7085 \pm 0.0518$ & $0.8786 \pm 0.0112$ & $\textbf{0.9236} \pm 0.0015$ \\
Accuracy & \dataset{} & $0.5632 \pm 0.0035$ & $0.7778 \pm 0.0690$ & $0.8963 \pm 0.0035$ & $\textbf{0.9315} \pm 0.0020$ \\
\midrule
Word & \texttt{BitCore} & $0.9759 \pm 0.0032$ & $0.9951 \pm 0.0007$ & $0.9976 \pm 0.0003$ & $\textbf{0.9986} \pm 0.0002$ \\
Level & \texttt{BitViper} & $0.4446 \pm 0.0003$ & $0.6192 \pm 0.0576$ & $0.8242 \pm 0.0150$ & $\textbf{0.8862} \pm 0.0021$ \\
Jaccard & \dataset{} & $0.4861 \pm 0.0018$ & $0.7041 \pm 0.0768$ & $0.8491 \pm 0.0048$ & $\textbf{0.8979} \pm 0.0024$ \\
\midrule
 & \texttt{BitCore} & $0.9923 \pm 0.0013$ & $0.9984 \pm 0.0003$ & $0.9992 \pm 0.0001$ & $\textbf{0.9996} \pm 0.0001$ \\
BLEU & \texttt{BitViper} & $0.7624 \pm 0.0003$ & $0.8563 \pm 0.0259$ & $0.9399 \pm 0.0053$ & $\textbf{0.9618} \pm 0.0007$ \\
 & \dataset{} & $0.7803 \pm 0.0009$ & $0.8907 \pm 0.0344$ & $0.9485 \pm 0.0017$ & $\textbf{0.9656} \pm 0.0009$ \\
\bottomrule
\end{tabularx}
\caption{\label{tab:comp_crossval}Comparison results of Character BERT-based restoration in terms of three evaluation measures with different amounts of training set}
\end{table*}

\begin{table}[!t]
\centering
\begin{tabular}{l ccc}
\toprule
\multicolumn{1}{c}{\multirow{2}{*}{Dataset}} & Original & Ground & Restored \\
 & VP word & Truth & Word \\
\midrule
% \texttt{BitCore} & on\g{\unicode{013a}}ine & online & on\ul{i}ine \\
 & \g{\unicode{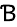}}\g{\unicode{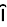}}tcoin & bitcoin & \ul{r}\ul{k}tcoin \\
\multirow{2}{*}{\texttt{BitCore}} & \g{\unicode{0181}}\g{\unicode{020b}}\g{\unicode{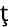}}\g{\unicode{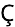}}oin & bitcoin & \ul{a}\ul{k}\ul{o}\ul{m}oin \\
 & \g{\unicode{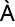}}\g{\unicode{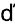}}\g{\unicode{010f}}res\g{\unicode{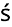}} & address & \ul{u}ddress \\
 & g\g{\unicode{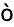}}og & good & goo\ul{g} \\
\midrule
 & \g{\unicode{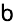}}ec\g{\unicode{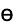}}me & became & bec\ul{o}me \\
\multirow{2}{*}{\texttt{BitViper}} & br\g{\unicode{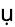}}\g{\unicode{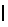}}e & brute & br\ul{o}\ul{k}e \\
 & \g{\unicode{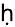}}\g{\unicode{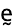}}a\g{\unicode{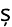}}t & beast & \ul{e}east \\
 & selle\g{\unicode{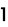}} & seller & selle\ul{t} \\
%  & \g{\unicode{1e2b}}\g{\unicode{04d9}}ve & have & \ul{i}\ul{\textquotesingle }ve \\
\midrule
 & tr\g{\unicode{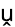}}l\g{\unicode{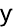}} & truly & trul\ul{l} \\
\multirow{2}{*}{\dataset{}} & \g{\unicode{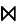}}all & mail & ma\ul{l}l \\
 & se\g{\unicode{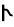}}\g{\unicode{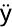}}i\g{\unicode{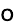}}e & service & se\ul{n}\ul{s}i\ul{d}e \\
%  & ad\g{\unicode{1e7d}}i\g{\unicode{1ecd}}e & advice & advi\ul{r}e \\
 & breako\g{\unicode{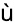}}\g{\unicode{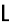}} & breakout & breakou\ul{s} \\
\bottomrule
\end{tabular}
\caption{\label{tab:incorBERT} List of VP words in three datasets that are incorrectly restored by the Character BERT-based methods from the experiments of Table~\ref{tab:comp_three} (VP characters are highlighted in gray color.)}
\end{table}

% Figure~\ref{fig:vpc_histogram} shows the histogram of the number of VP sentences according to the occurrence ratio of VP characters against the length of the sentence in \texttt{BitCore}, \texttt{BitViper}, and \dataset{} datasets, respectively. The x- and y-axes of each figure represent the ratio of VP characters included and the number of corresponding VP sentences, respectively. As shown in Figure~\ref{fig:vpc_histogram}(a), the VP sentences collected from bitcoinabuse[.]com do not yield unimodal distribution regarding the number of VP characters included. Rather, it has three peaks regarding the VP character ratio, such as 0.07 to 0.09, 0.32 to 0.34, and 0.66 to 0.68, that may be useful for devising VP restoration methods. Figure~\ref{fig:vpc_histogram}(b) representing the histogram of \texttt{BitViper} dataset indicates that its distribution of significantly different to that of \texttt{BitCore} dataset. Figure~\ref{fig:vpc_histogram}(c) shows the histogram of \dataset{} dataset. Table~\ref{tab:histogram_peak_examples} lists VP sentence examples of the three peaks in the histogram of \texttt{BitCore} dataset shown in Figure~\ref{fig:vpc_histogram}(a).

Next, the number of VP sentences according to the occurrence ratio of VP characters to sentence length in the \texttt{BitCore}, \texttt{BitViper}, and \dataset{} datasets is presented as a histogram as in Figure~\ref{fig:vpc_histogram} in Appendix \ref{sec:bitabuse_statistics}. Figure~\ref{fig:bitcore_clustering} shows the VP character-word association graph using the Yifan Hu algorithm \citep{hu2005efficient} in Gephi \citep{bastian2009gephi}. This graph represents the clustering of the VP character-word association of \texttt{BitCore} dataset, where nodes correspond to characters and words subjected to perturbation attacks. The distance between nodes indicates the degree of their relatedness\footnote{The Yifan Hu algorithm uses a multiscale approach to position highly related nodes close to each other while placing less related nodes further apart. This algorithm is fundamentally based on a force-directed layout, where nodes are arranged according to the forces of attraction and repulsion between them based on the frequency of association.}. Figure~\ref{fig:bitcore_clustering}(a) illustrates the overall graph, and Figures~\ref{fig:bitcore_clustering}(b)--(e) each represents the graph regarding key characters. Notably, the core of the major clusters is occupied by vowels such as `a', `e', `i', and `o'. This likely occurs because vowels are frequently used across various words, resulting in strong associations within the graph and positioning them at the center. Specifically, the character `e' appears to play a more global role within the graph, whereas other characters show stronger relations with words belonging to different clusters.

Table~\ref{tab:comp_three} shows the restoration performance of SimChar DB, OCR, Spell Checker, Character BERT, and GPT-4o mini-based methods on three datasets.
Experimental results indicate that the Character BERT-based method significantly outperforms the other three methods.
Regarding each dataset, all five methods achieved the best and worst performance for \texttt{BitCore} and \texttt{BitViper} datasets, respectively.
Table~\ref{tab:result_restore} represents two examples of restoration results regarding five methods. Although the Character BERT-based method restores two VP sentence examples perfectly, Table~\ref{tab:comp_three} indicates that the restoration performance of Character BERT is imperfect. Table~\ref{tab:incorBERT} lists VP words in three datasets that are incorrectly restored by the Character BERT-based method. The table shows that it often fails to restore if two or more VP characters are continued in the corresponding VP word.

We evaluated the Word Level Accuracy regarding the proportion of VP characters in sentences to validate the robustness of each method, as shown in Figure~\ref{fig:homoglyph_ratio_acc}.
In this experiment, the Character BERT-based method showed robust performance on both \texttt{BitCore} and \dataset{} datasets. It is interesting to note that it loses its robustness on \texttt{BitViper} dataset that does not include \texttt{BitCore} dataset, indicating that \texttt{BitCore} significantly contributes to the robust performance of the Character BERT-based method.
In summary, the Character BERT-based method showed the most robust performance for VP sentences with a high VP character ratio.
To see experimental results regarding Word Level Jaccard and BLEU, please refer to Appendix~\ref{sec:vpratio}.

Test performance with VP characters unseen during the training phase can be critical for the Character BERT-based method. Additional experiments were conducted using the Character BERT-based method with varying amounts of training VP sentences to validate this aspect. Specifically, when the amount of training VP sentences is extremely small, the Character BERT-based method encounters many unseen VP characters. Table~\ref{tab:comp_crossval} presents the performance of the Character BERT-based method when the proportion of training VP sentences is set to 1\%, 5\%, 10\%, and 20\%, respectively. In these experiments, we recognized that differences in the size of the test dataset could impact the fairness of performance comparisons. To address this, we sampled the remaining data, which were not used for training or validation, to standardize the size of the test dataset. Performance evaluation was conducted by measuring the performance of each pattern in the test dataset and calculating the mean and variance. This process was grounded in the Law of Large Numbers to include as many samples as possible, aiming to approximate the population mean.

The experimental results revealed that when the amount of training VP sentences was as low as 1\% or 5\%, significant performance degradation was observed for both the \texttt{BitViper} and \dataset{} datasets. This finding suggests that sufficient VP sentences are necessary to build a stable Character BERT model for VP text restoration. On the other hand, despite the performance drop with lower proportions of training data, the Word Level Accuracy still exceeded 0.5. This indicates that the model can restore relatively well from unseen attacks, even when it is exposed to many new VP attacks during the test phase. Additionally, using a smaller amount of training data allows the model to complete training more quickly, which is a desirable attribute in practical applications.

\section{Discussion}
\label{sec:discuss}

Figure~\ref{fig:homoglyph_ratio_acc} demonstrates the performance of various restoration methods based on the proportion of VP characters. Compared to other methods, the Character BERT-based method performed more effectively as the proportion of VP characters increased. This indicates that the Character BERT-based method can accurately restore VP characters by leveraging contextual information. Conversely, the Spell Checker-based method exhibited a sharp decline in performance as the proportion of VP characters increased, highlighting the limitations of simply correcting typographical errors when dealing with text containing a high proportion of VP characters.
The GPT-4o mini-based method underperformed compared to the spell checker-based method, likely because the GPT-4o mini is a generative model. Consequently, the word order and indexing between the input and output sentences are not maintained. This trait of generative language models seems to result in reduced performance in word accuracy evaluations, where word positioning is critical. Furthermore, the performance decline caused by the refusal responses triggered by the safety features of the language model, which will be discussed later, is also thought to have contributed to these results.
Figures~\ref{fig:wlj_ratio} and~\ref{fig:bleu_ratio}, like Figure~\ref{fig:homoglyph_ratio_acc}, evaluate the Word Level Jaccard and BLEU performance for the ratio of VP characters in sentences. The Jaccard performance closely mirrored the Word Level Accuracy results, but the BLEU performance exhibited a slightly different pattern. In both Figures~\ref{fig:homoglyph_ratio_acc} and~\ref{fig:wlj_ratio}, the Character BERT-based method consistently demonstrated the most effective performance as the VP character ratio increased, with significant performance gaps between it and the other methods. However, in Figure~\ref{fig:bleu_ratio}, the performance of all methods, except for the Character BERT-based method, generally improved, reducing the performance gap. This suggests that the BLEU score is more sensitive to contextual accuracy, meaning that even if the exact words do not match, simpler methods can achieve higher scores as long as the sentence structure and meaning are somewhat preserved.

% Table~\ref{tab:result_word_acc} shows the Word Level Accuracy of the four restoration methods, emphasizing that the Character BERT-based method demonstrated superior performance on the PhishiRest dataset. This suggests that the Character BERT-based method can also effectively operate in real phishing attack scenarios. The outstanding performance on the PhishiRest dataset indicates that our method can be a useful tool in practical environments.

Table~\ref{tab:comp_three} presents the comparison results of five restoration methods across three datasets using three evaluation measures. 
The results show that the Character BERT-based method clearly outperformed the others, with all approaches achieving the highest performance on the \texttt{BitCore} dataset and the lowest performance on the \texttt{BitViper} dataset.
Examples of the restoration for VP sentences through each method are shown in Table~\ref{tab:result_restore}. Although the Character BERT, GPT-4o mini, and Spell Checker share the commonality of leveraging contextual information, the character BERT-based method was more accurate in the restoration.
The SimChar DB-based method could only restore VP characters included in SimChar DB, and many of the VP characters that appeared did not exist in the DB, resulting in poor restoration performance.
Additionally, a fundamental limitation of simple mapping-based methods like SimChar DB is their inability to handle one-to-many mappings for VP characters. Since these methods are rule-based, they can only output a single non-VP character for each VP character. We will demonstrate how frequently one-to-many corresponding VP characters appear in the dataset in Appendix~\ref{sec:o2m_vp_chars}.
The OCR-based method also had poor restoration capability for each VP character, and it was observed that character recognition was more difficult in the case of VP characters containing diacritics. The Spell Checker-based method showed high performance in restoring words containing VP characters, but it occasionally failed to find suitable words when the VP character ratio in the sentence was high. 
The GPT-4o mini-based method showed limited restoration capabilities. While it was able to successfully restore most VP characters in cases like Example 1, it failed when VP characters dominated the sentence, as in Example 2, producing outputs that differed significantly from the input. Additionally, in certain cases, due to the language model’s safety features, responses such as ``I'm sorry, but I can't assist with that,'' ``I'm sorry, I can't assist with that,'' or ``I'm sorry, I can't help with that'' were generated in response to unethical content. These instances made up about 13.22\% of the \dataset{}, which is a notable proportion.
The Character BERT-based method excels by directly learning the context and succeeded in almost perfectly restoring VP words. This implies that models like BERT, which are significantly smaller than generative large language models such as GPT-4, can be more efficient for restoring VP text, as they still achieve high performance despite their smaller size.
Table~\ref{tab:incorBERT} provides examples of VP words that were incorrectly restored using the Character BERT-based method. The results indicate that restoration failures are more likely when VP characters appear consecutively or when there is a high density of attacked VP characters nearby. 
% Additionally, there is a tendency for restoration errors when VP characters are restored to multiple types of alphabetic characters.

As demonstrated by the comparison results in Table~\ref{tab:comp_three}, the Character BERT-based method achieved nearly 100\% accuracy on \texttt{BitCore} dataset, highlighting its robustness and reliability. In addition, with sufficient training VP sentences, it achieved almost perfect performance on \dataset{} dataset as shown in Table~\ref{tab:comp_crossval}.
% 아래 내용은 데드라인 걸려서 못 넣음.
% While the \texttt{BitCore} dataset achieves high accuracy with only about 300 training samples (1\% of the total), BitViper shows low accuracy even with 2,000 samples (also 1\%). This suggests: 1) BitViper generates data that is overly difficult compared to real-world scenarios, 2) in reality, high restoration accuracy can be achieved with less data, making Character BERT highly effective for VP character restoration, and 3) language models excel at defending against real-world phishing attacks, showing their potential in security.

Given the high performance of LM pre-trained using \dataset{}, it may be employed in highly specialized, high-performance restoration tasks. For example, the pre-trained model could be applied in digital forensics to decode and reconstruct documents, emails, or logs that have been intentionally manipulated to obscure evidence. In addition, the model can be further trained to effectively handle even subtle and complex text modifications, which could improve forensic analysis. We believe that this model could also be used in secure messaging systems, where it would restore the original content of messages that have been deliberately obfuscated to ensure the secure transmission of sensitive information. These studies may highlight the potential of our datasets and pre-trained models to address critical challenges in secure communications.

\section{Conclusion}

In this study, we created three VP text datasets: \texttt{BitCore}, \texttt{BitViper}, and \dataset{}. Our analysis results show that \texttt{BitCore} and \texttt{BitViper} have significantly different characteristics, and the LM-based reconstruction method demonstrates strong robustness and potential on all three datasets. \dataset{}, a pre-trained model using 325,580 VP sentences, can be downloaded from \dataset{}.\footnote{\url{https://huggingface.co/datasets/AutoML/bitaubse}}. In future studies, a hybrid approach, such as combining OCR and Character BERT, can be explored to achieve robust performance with insufficient training samples. Internalizing them into LMs may be beneficial for remedying the greedy data consumption nature of LMs and in scenarios where collecting sufficient samples is challenging. In addition, lightweight yet accurate LMs for restoration tasks may be obtained if the bias to the words attacked frequently and vowel characters in real-world phishing attacks is exploited effectively. Lastly, validating the zero-shot performance of \dataset{} model should also be performed.

\clearpage

\section*{Limitations}

The VP text restoration experiments conducted in this study did not include additional restoration methods to avoid exceeding the scope of the study. Specifically, a performance comparison between the Character BERT-based and other LM-based restoration methods was not performed. Thus, it is difficult to evaluate the superiority of Character BERT over other modern LMs. Character BERT showed sufficiently good performance, but it will be possible to compare effectiveness and efficiency with methods applying other LMs in the future.

The \dataset{} dataset used in this study only includes data related to Bitcoin scams, which limits its ability to reflect a variety of phishing attack scenarios. In addition, phishing attacks may appear in more diverse or complex forms over time, and failure to reflect this diversity may reduce the generalizability of our study. Thus, future studies should aim to construct an extended dataset that includes various phishing attack scenarios and conduct studies comparing different restoration methods.

Also, our datasets were created for study purposes to defend against phishing attacks based on VP texts. However, there is a risk that this dataset could be used by non-experts in phishing to learn and execute attacks.} For example, WormGPT, recently created on the dark web to generate criminal text, and PoisonGPT, released by Mithril Security, spread contaminated results. These models might use our datasets to develop malicious tools. Consequently, this could lead to the sophistication of phishing attacks, resulting in more victims. In addition, the damage caused by the misuse of such datasets is difficult to hold accountable legally. Currently, many countries lack clear regulations regarding the technological misuse of such datasets, necessitating careful considerations and observations. The datasets and models used in this paper are publicly available, but they should not be used for purposes other than research.

\section*{Ethics Statement}

Our datasets were created for study purposes to defend against phishing attacks based on VP texts. However, there is a risk that this dataset could be used by non-experts in phishing to learn and execute attacks. For example, WormGPT, recently created on the dark web to generate criminal text, and PoisonGPT, released by Mithril Security, spread contaminated results. These models might use our datasets to develop malicious tools. Consequently, this could lead to the sophistication of phishing attacks, resulting in more victims. In addition, the damage caused by the misuse of such datasets is difficult to hold accountable legally. Currently, many countries lack clear regulations regarding the technological misuse of such datasets, necessitating careful considerations and observations. The datasets and models used in this paper are publicly available, but they should not be used for purposes other than research.

\section*{Acknowledgements}

This research was supported by the Chung-Ang University Graduate Research Scholarship in 2024 and by the Institute of Information \& Communications Technology Planning \& Evaluation (IITP) grant funded by the Korea government (MSIT) (2021-0-01341, Artificial Intelligence Graduate School Program (Chung-Ang University)).

This work was supported by Institute of Information \& communications Technology Planning \& Evaluation (IITP) grant funded by the Korea government (MSIT) (No. RS-2024-00402898, Simulation-based High-speed/High-Accuracy Data Center Workload/System Analysis Platform)

\bibliography{custom}
\bibliographystyle{acl_natbib}

\clearpage

\begin{table*}[!t]
    \centering
    \begin{minipage}[t]{.48\linewidth}
        \centering
        \begin{tabular}{lr}
        \toprule
        \multicolumn{1}{c}{Category} & \multicolumn{1}{c}{Value}\\
        \midrule
        Number of email texts & 262,258\\
        Min. length of email text & 10\\
        Max. length of email text & 2,000\\
        Average length of email text & 417 \\
        \bottomrule
        \end{tabular}
        \caption{\label{tab:raw_dataset}Brief statistics of phishing emails collected from bitcoinabuse[.]com}
    \end{minipage}
    \begin{minipage}[t]{.48\linewidth}
        \centering
        \begin{tabular}{lr}
        \toprule
        \multicolumn{1}{c}{Category} & \multicolumn{1}{c}{Value}\\
        \midrule
        Number of sentences & 325,580\\
        Number of VP sentences & 26,591\\
        Number of non-VP sentences & 298,989\\
        Average length of sentences & 91 \\
        \bottomrule
        \end{tabular}
        \caption{\label{tab:phishirest_dataset}Brief statistics of the raw corpus}
    \end{minipage}
\end{table*}

% <정규표현식 예시 표>
\begin{table*}[!t]
\centering
{
\newcolumntype{L}{>{\raggedright\arraybackslash}X}
\scriptsize
\begin{tabularx}{\textwidth}{LLl LLl}
\toprule
Description & Regular Expression & & Description & Regular Expression & \\
\midrule
Miscellaneous symbols & \verb|[\u 260e\u 2610\u 2611| \verb|\u 261e\u 2620\u 2639| \verb|\u 2640\u 2642\u 2661| \verb|\u 2665\u 267b\u 26a0| \verb|\u 26d4]| & R & Dingbats & \verb|[\u 2705\u 270a\u 270c| \verb|\u 270d\u 2714\u 2757| \verb|\u 2764\u 2795\u 2797| \verb|\u 27a1]| & R \\
General punctuations and formatting characters & \verb|[\u 200b-\u 200d\u 2022| \verb|\u 202a\u 2028\u 2039| \verb|\u 203a\u 2060-\u 2069]| & R & Emoticons, HTML tag patterns, and special character sequences & ¯\verb|\\_\(|\begin{CJK}{UTF8}{min}ツ\end{CJK}\verb|\)_\/|¯ | \& \verb|#8203;| | </?sp.n>| \textbackslash?\verb|\ u200d| [\female\mars] | \verb|\*| †† \verb|\*| \verb|\*| & R \\
Latin supplements & \verb|[\u 00a7\u 00a9\u 00ab-| \verb|\u 00ae\u 00b0\u 00b7| \verb|\u 00bb\u 00bf]| & R & Control characters & \verb|[\u 0000\u 0006-\u 0008| \verb|\u 000b-\u 001f\u 0080-| \verb|\u009f]| & R \\
Bitcoin wallet address & \verb|[13][a-km-zA-HJ-NP-| \verb|Z1-9]{25,34}| & R & Email address & \verb|[\w-\.]+@([\w-]+\.)+| \verb|[\w-]{2,4}| & R \\
CJK characters & \verb|[\u 3040-\u 9fff\u ac00-| \verb|\u d7ff]| & R & Box elements / geometric shapes & \verb|[\u 2592\u 25a0\u 25cb| \verb|\u25cf]| & R \\
Emoji etc. & \verb|[\u 1f000-\u 1ffff]| & R & Private use area & \verb|[\u e000-\u f8ff]| & R \\
Variation selectors & \verb|[\u fe00-\u fe0f]| & R & Combining diacritical marks & \verb|[\u032a\u034f]| & R \\
Arabic characters & \verb|[\u 061c\u 0640]| & R & Sinhala characters & \verb|[\u 0d9a\u 0dd4]| & R \\
Letter-like symbols & \verb|[\u 2116\u 2122]| & R & Mathematical operators & \verb|[\u 2211\u 22ef]| & R \\
Miscellaneous symbols and arrows & \verb|[\u 2b07\u 2b55]| & R & Halfwidth and fullwidth forms & \verb|[\u ff0a\u ff5e]| & R \\
Modifier letter up arrowhead & \verb|[\u 02c4]| & R & Superscript six & \verb|[\u 2076]| & R \\
Combining enclosing keycap & \verb|[\u 20e3]| & R & Upwards arrow & \verb|[\u 2191]| & R \\
Top half integral symbol & \verb|[\u 2320]| & R & Zero width no-break space & \verb|[\u feff]| & R \\
Special space characters & \verb|[\u 00a0\u 2002-\u 200a| \verb|\u 3000]| & S & Small quotation mark, accent mark, or prime symbol & \verb|[\u00b4\u02bb\u02cb| \verb|\u 2018\u 2019\u 2032]| & \textquotesingle \\
Diaeresis, double quotation mark, or double prime symbol & \verb|[\u 00a8\u 201c\u 201d| \verb|\u 2033\u 275d\u 275e]| & \textquotedbl & Various types of hyphens, dashes, or the minus sign & \verb|[\u 2010\u 2011\u 2013| \verb|\u 2014\u 2015\u 2212]| & - \\
Low quotation mark or a fullwidth comma & \verb|[\u 201a\u 201e\u ff0c]| & , & Double exclamation mark or a fullwidth exclamation mark & \verb|[\u 203c\u ff01]| & ! \\
Various types of left brackets & \verb|[\u 300a\u 3010\u ff08]| & ( & Various types of right brackets & \verb|[\u 300b\u 3011\u ff09]| & ) \\
Various types of equals sign & \verb|[\u 2248\u ff1d]| & = & Horizontal ellipsis & \verb|[\u 2026]| & ... \\
Fullwidth colon & \verb|[\u ff1a]| & : &  Text decoding errors & \euro \texttrademark & \textquotesingle\\
Fullwidth semicolon & \verb|[\u ff1b]| & ; & &\^a \euro [ \oe ] \verb|?|& \textquotedbl\\
Multiplication sign & \verb|[\u 00d7]| & x \\
\bottomrule
\end{tabularx}
}
\caption{Preprocessed characters represented in their Unicode based on corresponding regular expressions}
\label{tab:regex_examples}
\end{table*}

\begin{table*}[!t]
\centering
\small
{
\newcolumntype{L}{>{\raggedright\arraybackslash}X}
\begin{tabularx}{\textwidth}{L L}
\toprule
\multicolumn{1}{c}{Original Text} & \multicolumn{1}{c}{Preprocessed Text} \\
\midrule
\unicode{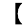} Reminder \unicode{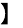} Your system devices has been Hacked \unicode{3010} National Security Agency \unicode{3011} Authority-11622272 & ( Reminder ) Your system devices has been Hacked ( National Security Agency ) Authority-11622272 \\
\midrule
Af\ZWS te\ZWS r \ZWS re\ZWS ce\ZWS iv\ZWS in\ZWS g \ZWS th\ZWS e \ZWS pa\ZWS ym\ZWS en\ZWS t,\ZWS\ I\ZWS\ w\ZWS il\ZWS l \ZWS de\ZWS le\ZWS te\ZWS\ t\ZWS he\ZWS\ v\ZWS id\ZWS eo \ZWS , & After receiving the payment, I will delete the video,\\
\midrule
You may not know me \unicode{0430}\&\#8203;nd \unicode{0443}\&\#8203;ou are pr\unicode{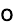}\&\#8203;b\unicode{0430}\&\#8203;bly\&\#8203; & You may not know me \unicode{0430}nd \unicode{0443}ou are pr\unicode{043e}b\unicode{0430}bly \\
\bottomrule
\end{tabularx}
}
\caption{Example of text preprocessed using regular expressions. The red box with the number in it indicates the unprintable Unicode character of the hex value written inside it (Please see color PDF.)} 
\label{tab:replace_example}
\end{table*}

\begin{table*}[!t]
\centering
{
\newcolumntype{L}{>{\raggedright\arraybackslash}X}
\begin{tabularx}{\textwidth}{L rrrrrr}
\toprule
\multirow{2}{*}{Dataset} & \multicolumn{1}{c}{Number of} & \multicolumn{1}{c}{Average} & \multicolumn{1}{c}{Number of} & \multicolumn{1}{c}{Unique VP} & \multicolumn{1}{c}{Number of} & \multicolumn{1}{c}{Unique VP} \\
& \multicolumn{1}{c}{VP Sentences} & \multicolumn{1}{c}{Length} & \multicolumn{1}{c}{VP Words (\%)} & \multicolumn{1}{c}{Words} & \multicolumn{1}{c}{VP Characters (\%)} & \multicolumn{1}{c}{Characters} \\
\midrule
\texttt{BitCore} & 26,591 & 92 & 261,460 (58\%) & 37,726 & 503,239 (26\%) & 317 \\
\texttt{BitViper} & 298,989 & 91 & 2,861,434 (58\%) & 1,126,986 & 4,347,988 (20\%) & 525 \\
\dataset{} & 325,580 & 91 & 3,122,894 (58\%) & 1,160,211 & 4,851,227 (21\%) & 706 \\
\bottomrule
\end{tabularx}
}
\caption{Brief statistics of \texttt{BitCore}, \texttt{BitViper}, and \dataset{} datasets}
\label{tab:datasets_stat}
\end{table*}

\clearpage
\appendix

\begin{table}[!t]
\centering
{
\newcolumntype{L}{>{\raggedright\arraybackslash}X}
\newcolumntype{R}{>{\raggedleft\arraybackslash}X}
\small
\begin{tabularx}{\columnwidth}{cR | cR}
\toprule
\multirow{2}{*}{Word} & \multicolumn{1}{c|}{Number of} & \multirow{2}{*}{Word} & \multicolumn{1}{c}{Number of} \\
 & \multicolumn{1}{c|}{Variants} &  & \multicolumn{1}{c}{VP attacked} \\
\midrule
your & 369 & you & 15,103\\
access & 361 & your & 10,725\\
email & 293 & to & 7,745\\
software & 286 & and & 7,626\\
Bitcoin & 268 & the & 6,906\\
videos & 266 & a & 5,277\\
video & 265 & I & 4,005\\
have & 254 & have & 3,781\\
transfer & 230 & this & 3,685\\
bitcoin & 227 & video & 3,576\\
internet & 225 & that & 3,103\\
browsing & 220 & of & 2,881\\
you & 207 & know & 2,713\\
which & 205 & will & 2,436\\
about & 200 & is & 2,407\\
will & 198 & all & 2,391\\
contacts & 195 & on & 2,196\\
activities & 191 & what & 2,163\\
relatives & 187 & contacts & 1,850\\
with & 185 & as & 1,794\\
social & 173 & with & 1,779\\
devices & 171 & it & 1,773\\
account & 158 & i & 1,758\\
antivirus & 156 & software & 1,674\\
tracking & 155 & email & 1,654\\
watching & 154 & after & 1,625\\
after & 143 & from & 1,614\\
managed & 142 & access & 1,510\\
know & 137 & part & 1,430\\
from & 135 & site & 1,365\\
also & 134 & videos & 1,345\\
considering & 134 & in & 1,247\\
virus & 134 & me & 1,234\\
microphone, & 131 & are & 1,233\\
deactivate & 130 & not & 1,222\\
information & 130 & bitcoin & 1,197\\
this & 129 & which & 1,186\\
accounts & 125 & do & 1,170\\
according & 124 & watching & 1,159\\
received, & 123 & visited & 1,148\\
away & 122 & payment & 1,131\\
websites & 122 & can & 1,119\\
masturbating & 120 & for & 1,086\\
purchased & 119 & malware & 1,056\\
gained & 118 & porn & 1,029\\
signatures & 118 & don't & 985\\
happen & 117 & account & 976\\
installed & 117 & right & 923\\
months & 117 & screen & 848\\
simple & 117 & about & 843\\
\bottomrule
\end{tabularx}
}
\caption{The top 50 list of VP word variants and VP attacks for each word appearing in the \dataset{} dataset.} 
\label{tab:extended_interest_words}
\end{table}

\begin{table}[!t]
\centering
{
\newcolumntype{L}{>{\raggedright\arraybackslash}X}
\begin{tabularx}{\columnwidth}{cL}
\toprule
Word & \multicolumn{1}{c}{VP words} \\
\midrule
your & y\g{\unicode{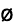}}ur, \g{\unicode{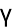}\unicode{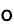}\unicode{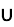}\unicode{0433}}, \g{\unicode{04af}}o\g{\unicode{03c5}\unicode{027e}}, \g{\unicode{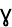}\unicode{043e}}u\g{\unicode{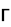}}, \g{\unicode{04af}\unicode{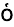}\unicode{1d1c}\unicode{1d26}}, \g{\unicode{04af}\unicode{1d0f}\unicode{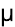}\unicode{1d26}}, y\g{\unicode{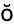}}ur, y\g{\unicode{0151}\unicode{0169}}r, \g{\unicode{04af}\unicode{043e}\unicode{1d1c}\unicode{0433}}, \g{\unicode{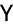}}0\g{\unicode{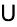}\unicode{053b}}, $\ldots$\\
access & \g{\unicode{03b1}}c\g{\unicode{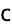}}e\g{\unicode{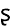}}s, \g{\unicode{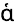}}cce\g{\unicode{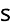}\unicode{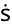}}, acc\g{\unicode{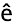}}ss, acce\g{\unicode{0455}\unicode{0455}}, acc\g{\unicode{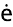}}ss, a\g{\unicode{0441}\unicode{0441}\unicode{0435}}ss, \g{\unicode{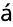}}c\g{\unicode{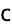}}e\g{\unicode{015b}\unicode{015b}}, $\ldots$\\
email & e\g{\unicode{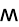}}a\g{\unicode{03b9}}l, \g{\unicode{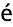}\unicode{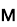}\unicode{0430}\unicode{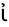}\unicode{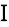}}, \g{\unicode{00e9}}m\g{\unicode{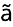}\unicode{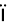}}l, \g{\unicode{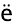}}ma\g{\unicode{00ef}}l, e\g{\unicode{1d0d}\unicode{0430}\unicode{00ed}}l, \g{\unicode{00ea}\unicode{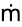}\unicode{0430}\unicode{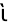}}l, e\g{\unicode{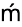}\unicode{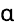}\unicode{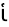}\unicode{04cf}}, \g{\unicode{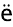}}ma\g{\unicode{00ed}\unicode{04cf}}, \g{\unicode{00e9}\unicode{1e3f}\unicode{00e1}\unicode{1f76}}l, $\ldots$\\
software & softw\g{\unicode{03b1}\unicode{1d26}}e, softw\g{\unicode{03b1}}re, softw\g{\unicode{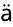}}re, s\g{\unicode{03c3}}ftw\g{\unicode{0251}\unicode{0433}\unicode{0435}}, \g{\unicode{0455}\unicode{1f41}\unicode{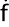}t\unicode{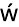}\unicode{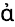}\unicode{0433}\unicode{00e9}}, s\g{\unicode{00f2}}ftwar\g{\unicode{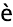}}, $\ldots$\\
Bitcoin & B\g{\unicode{1f30}}tc\g{\unicode{1d0f}\unicode{1f30}\unicode{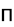}}, B\g{\unicode{00ed}}tco\g{\unicode{00ed}\unicode{043f}}, B\g{\unicode{1f31}}t\g{\unicode{0441}\unicode{03c3}\unicode{1f30}\unicode{1d28}}, \g{\unicode{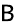}\unicode{03b9}}tc\g{\unicode{03bf}\unicode{03b9}\unicode{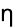}}, B\g{\unicode{1f31}}t\g{\unicode{0441}\unicode{03bf}\unicode{1f31}\unicode{1d28}}, 
\g{\unicode{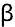}\unicode{03b9}}tc\g{\unicode{03bf}\unicode{03b9}\unicode{03b7}}, $\ldots$\\
\bottomrule
\end{tabularx}
}
\caption{\label{tab:vpw_example} Examples of VP variants regarding five words of Table~\ref{tab:extended_interest_words} with the highest number of variants (VP characters are highlighted in gray color.)}
\end{table}

\section{Statistics of Raw Dataset}
\label{sec:raw_dataset_statistics}

Table~\ref{tab:raw_dataset} shows brief statistics of the collected email texts. We identified 262,258 phishing-related emails from bitcoinabuse[.]com between May 16, 2017, and January 15, 2022, and extracted the text bodies of these emails. The length of the email bodies averages about 417 characters, ranging from a minimum of 10 characters to a maximum of 2,000 because the platform limits the maximum number of characters to 2,000. The content of phishing-related emails was uploaded from approximately 224 countries, and the country of upload and the language of the collected text may differ.

% <corpus 통계 표>
% \begin{table}[!t]
% \centering
% \begin{tabular}{lr}
% \toprule
% \multicolumn{1}{c}{Category} & \multicolumn{1}{c}{Value}\\
% \midrule
% Number of sentences & 325,580\\
% Number of VP sentences & 26,591\\
% Number of non-VP sentences & 298,989\\
% Average length of sentences & 91 \\
% \bottomrule
% \end{tabular}
% \caption{\label{tab:phishirest_dataset}\color{red}Brief statistics of the raw corpus}
% \end{table}

Table~\ref{tab:phishirest_dataset} presents the statistics of the raw corpus after splitting the collected texts into individual sentences and removing meaningless texts, as mentioned in the Data Collection section. The sentence-splitting process was performed using the NLTK library, resulting in a total of 325,580 sentences. In the next step, sentences containing non-ASCII characters were classified as VP sentences, and the classification was manually reviewed to ensure accuracy. After the review, 26,591 sentences were identified as VP sentences, while 298,989 were categorized as non-VP sentences.

\section{Filtering Non-English Texts}
\label{sec:filteringBERT}

\begin{table}[!t]
\centering
{
\newcolumntype{L}{>{\raggedright\arraybackslash}X}
\newcolumntype{R}{>{\raggedleft\arraybackslash}X}
\begin{tabularx}{\columnwidth}{cR | cR}
\toprule
\multirow{2}{*}{Char.} & \multicolumn{1}{c|}{Number of} & \multirow{2}{*}{Char.} & \multicolumn{1}{c}{Number of} \\
& \multicolumn{1}{c|}{Variants} & & \multicolumn{1}{c}{VP attacked} \\
\midrule
i & 34 & o & 179,792\\
a & 28 & i & 145,832\\
o & 27 & c & 95,863\\
e & 22 & a & 85,905\\
u & 22 & e & 57,009\\
r & 20 & n & 38,135\\
m & 14 & t & 38,085\\
n & 14 & d & 37,178\\
p & 13 & l & 34,865\\
s & 13 & u & 14,840\\
b & 13 & r & 14,381\\
t & 12 & s & 14,202\\
c & 12 & p & 13,717\\
w & 12 & v & 9,788\\
k & 12 & y & 9,073\\
y & 10 & h & 7,313\\
h & 9 & k & 5,966\\
l & 8 & m & 5,349\\
d & 8 & g & 3,452\\
j & 7 & f & 3,273\\
v & 6 & w & 2,855\\
g & 6 & b & 2,413\\
x & 5 & x & 480\\
q & 5 & q & 220\\
f & 3 & j & 107\\
z & 3 & z & 10\\
0 & 2 & 0 & 6\\
\bottomrule
\end{tabularx}
}
\caption{The full list of VP character variants and VP attacks for each character appearing in the \dataset{} dataset.} 
\label{tab:extended_interest_characters}
\end{table}

In our study, we exploited the BERT model with a fully connected classification layer trained to classify English texts from non-English texts. To train our model, we use the Flair library \citep{akbik2019flair}. In addition, the learning rate was set to $1e-6$, with 1 learning epoch (the library early stopped training due to the very small learning rate), a batch size of eight, an AdamW optimizer ($\beta_{1}=0.9$, $\beta_{2}=0.999$, $\textrm{weight\_decay}=0$), and the AnnealOnPlateau scheduler implemented in the Flair library. Additionally, a single NVIDIA GeForce RTX 3080 GPU was used.

\section{Regular Expressions}
\label{sec:regex_examples}

Table~\ref{tab:regex_examples} shows the list of regular expressions we used for further preprocessing. The first and fourth, the second and fifth, and the third and sixth columns mean the description of characters, regular expressions, and replaced characters, respectively. R and S in the third and sixth columns mean ``Removed'' and ``Space''. For example, No-break Space (U+00A0), En Space (U+2002), Hair Space (U+200A), and Ideographic Space (U+3000) are special space characters and would commonly be replaced with regular space characters. The space after \textbackslash{u} in the regular expression is included intentionally for clarity but is excluded in the actual regular expression. We also release a downloadable list of regular expressions and preprocessing code from \url{https://huggingface.co/datasets/AutoML/bitaubse/blob/main/preprocessing.py}.

% \begin{table*}[!t]
% \centering
% {
% \newcolumntype{L}{>{\raggedright\arraybackslash}X}
% \begin{tabularx}{\textwidth}{L L}
% \toprule
% \multicolumn{1}{c}{Original Text} & \multicolumn{1}{c}{Preprocessed Text} \\
% \midrule
% \unicode{3010} Reminder \unicode{3011} Your system devices has been Hacked \unicode{3010} National Security Agency \unicode{3011} Authority-11622272 & ( Reminder ) Your system devices has been Hacked ( National Security Agency ) Authority-11622272 \\
% \midrule
% Af\ZWS te\ZWS r \ZWS re\ZWS ce\ZWS iv\ZWS in\ZWS g \ZWS th\ZWS e \ZWS pa\ZWS ym\ZWS en\ZWS t,\ZWS\ I\ZWS\ w\ZWS il\ZWS l \ZWS de\ZWS le\ZWS te\ZWS\ t\ZWS he\ZWS\ v\ZWS id\ZWS eo \ZWS , & After receiving the payment, I will delete the video,\\
% \midrule
% You may not know me \unicode{0430}\&\#8203;nd \unicode{0443}\&\#8203;ou are pr\unicode{043e}\&\#8203;b\unicode{0430}\&\#8203;bly\&\#8203; & You may not know me \unicode{0430}nd \unicode{0443}ou are pr\unicode{043e}b\unicode{0430}bly \\
% \bottomrule
% \end{tabularx}
% }
% \caption{\color{red}Example of text preprocessed using regular expressions. The red box with the number in it indicates the unprintable Unicode character of the hex value written inside it (Please see color PDF.)} 
% \label{tab:replace_example}
% \end{table*}

Table~\ref{tab:replace_example} shows example sentences after the preprocessing based on the regular expressions. In three examples of the table, emojis and special characters in the sentence are removed, and unusual characters are replaced with ASCII characters with the same meaning. For example, in the first example in the table, ``Left Black Lenticular Bracket (U+3010)'' and ``Right Black Lenticular Bracket (U+3011)'' were replaced with regular parentheses (U+0028, U+0029). In the second example, unprintable Unicode characters that are presented as a hex value in the red box are removed.

\section{Statistics of \dataset{}}
\label{sec:bitabuse_statistics}

%table 6
We created \texttt{BitCore}, \texttt{BitViper}, and \texttt{BitAbuse} datasets based on the raw corpus. Brief statistics of the three datasets are presented in Table~\ref{tab:datasets_stat}. Specifically, \texttt{BitCore} was created by simply selecting 26,591 VP sentences from the raw corpus. Next, \texttt{BitViper} was created by applying the character perturbation procedure of Viper that uses the ICEs method with a probability of 0.2 to 298,989 non-VP sentences of the raw corpus, following the same settings used in the original study for the restoration task\footnote{TextBugger is not considered here because it attacks by altering keywords in sentences for semantic classification. Thus, applying TextBugger to non-VP texts of the raw corpus requires additional work, such as labeling whether a sentence is spam, which is out of the scope of this study.}. Lastly, \dataset{} was created by merging \texttt{BitCore} and \texttt{BitViper}, resulting in the largest dataset of our study that contains both real-world and synthetic VP sentences.

\section{Statistics and Examples of VP Words and Characters}
% \label{sec:extended_lists}

\begin{figure*}
    \centering
    \begin{subfigure}[b]{0.30\textwidth}
        \centering
        \includegraphics[width=\textwidth]{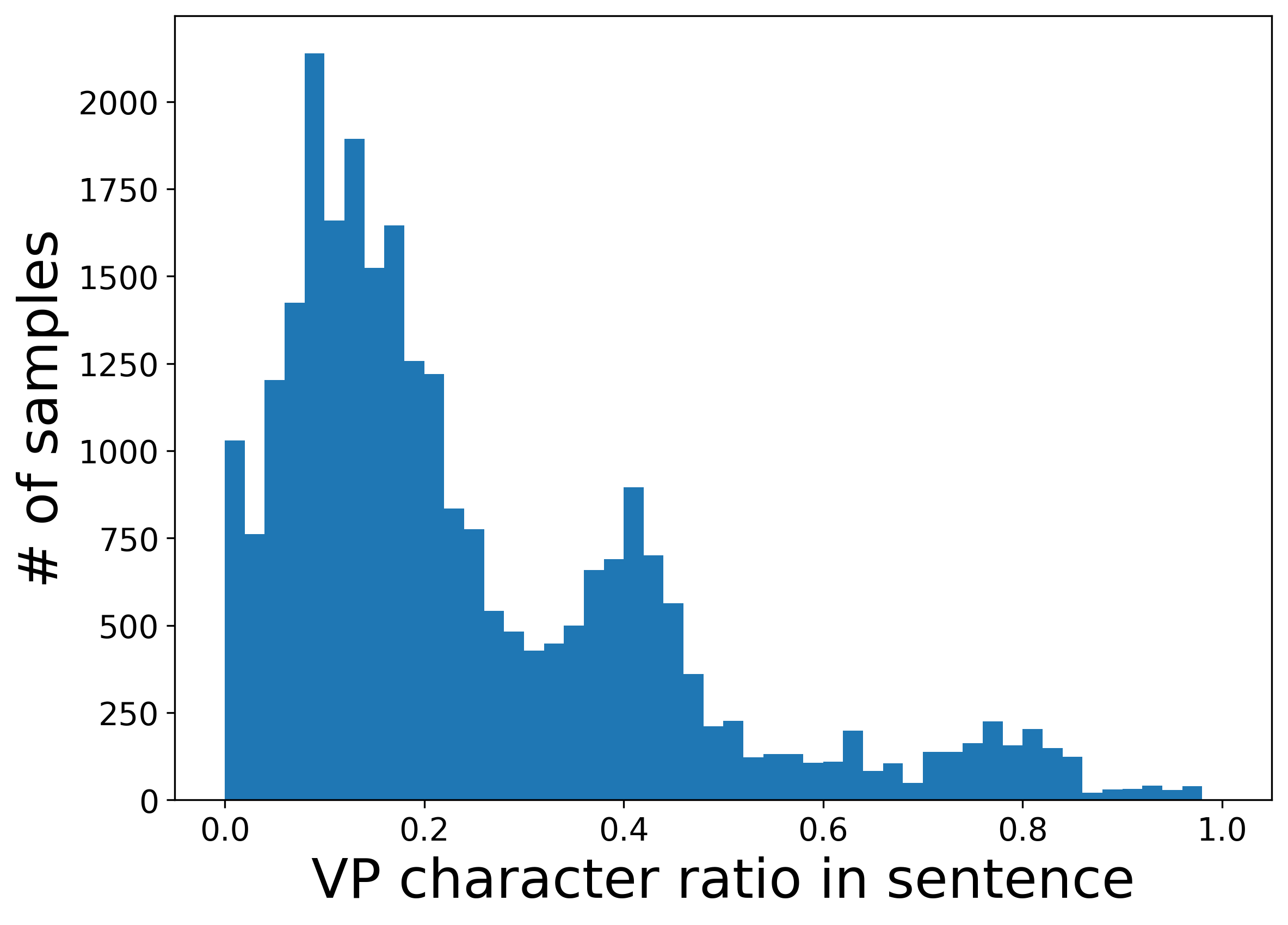}
        \caption{\texttt{BitCore} dataset (Real)}
        \label{fig:phishirest_histogram}
    \end{subfigure}
    \hfill
    \begin{subfigure}[b]{0.30\textwidth}
        \centering
        \includegraphics[width=\textwidth]{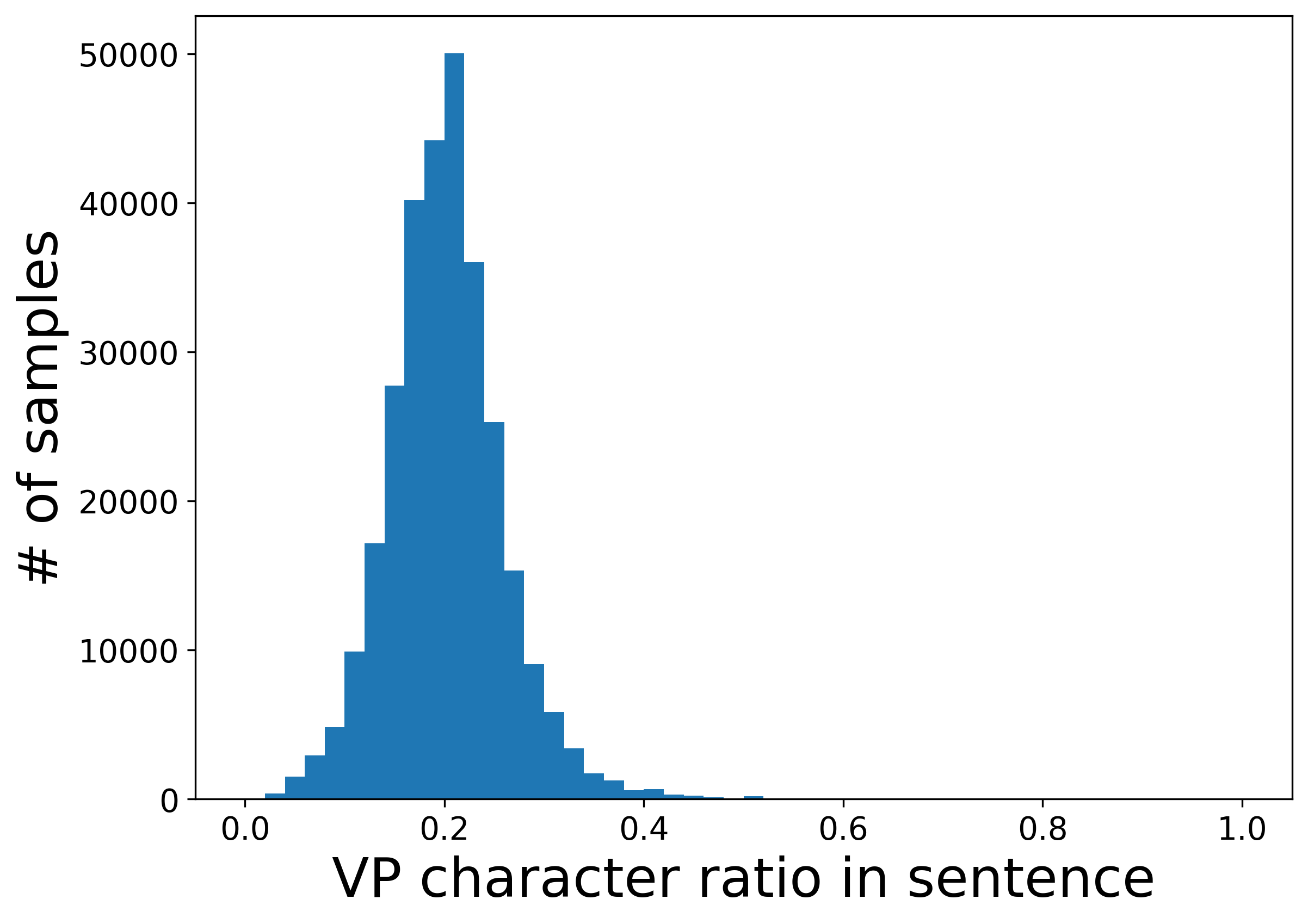}
        \caption{\texttt{BitViper} dataset (Synthesized)}
        \label{fig:phishirest_viper_histogram}        
    \end{subfigure}
    \hfill
    \begin{subfigure}[b]{0.30\textwidth}
        \centering
        \includegraphics[width=\textwidth]{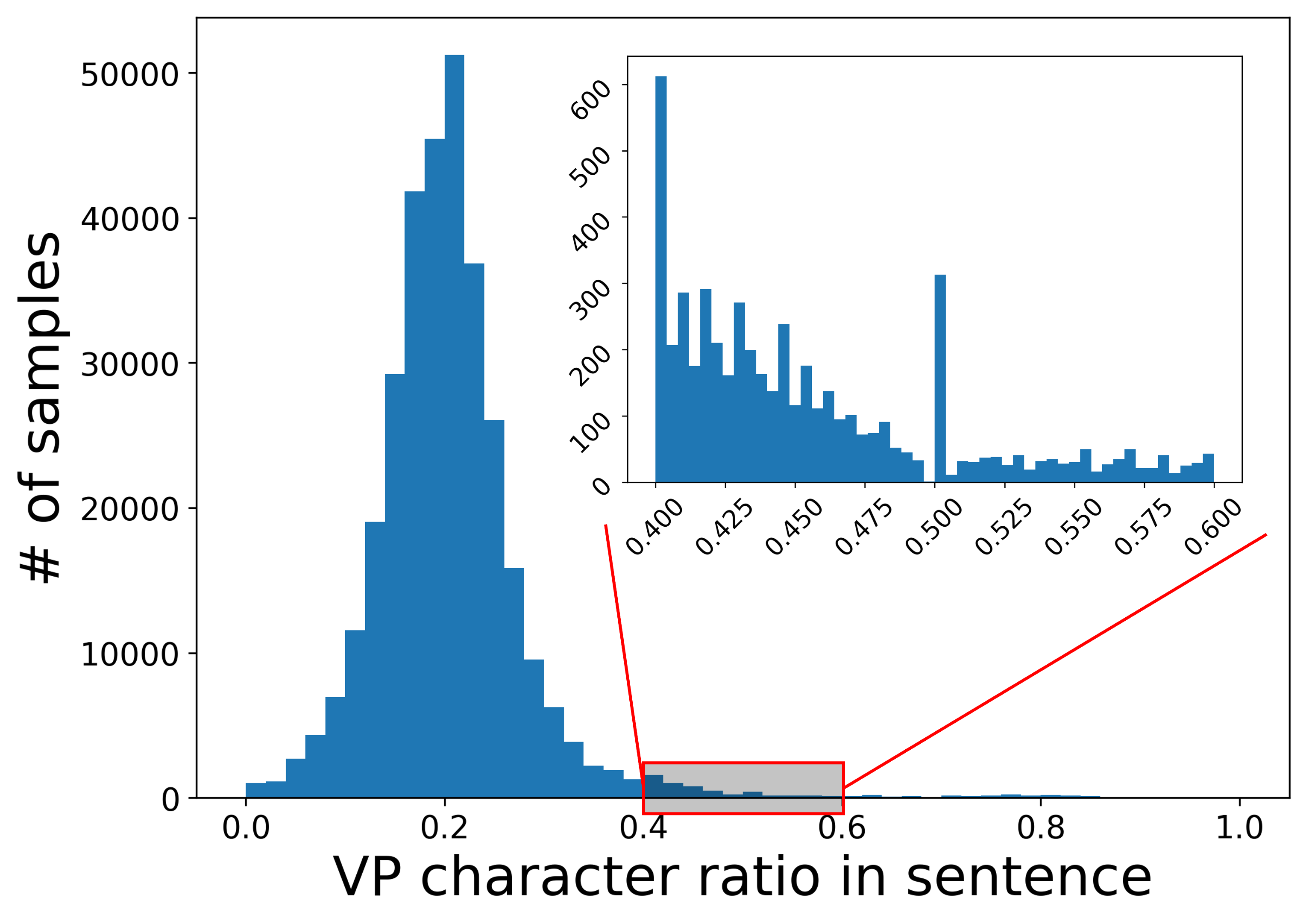}
        \caption{\dataset{} dataset (Mixed)}
        \label{fig:bitabuse_histogram}
    \end{subfigure}
\caption{\label{fig:vpc_histogram} The histogram of the number of VP sentences \texttt{BitCore} (26,591 sentences), \texttt{BitViper} (298,989 sentences), and \dataset{} datasets (325,580 sentences) according to the occurrence ratio against the sentence length.}
\end{figure*}

\begin{table*}[!t]
\centering
{
\newcolumntype{L}{>{\raggedright\arraybackslash}X}
\begin{tabularx}{\textwidth}{lL}
\toprule
Peak & VP sentence examples \\
\midrule
0.07 - 0.09 & if\ you\ do\ n\g{\unicode{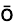}}t\ fund\ th\g{\unicode{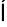}}s\ bitc\g{\unicode{014d}}in\ address\ with\ \$1000\ with\g{\unicode{013a}}n\ n\g{\unicode{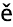}}xt\ 2\ days,\ i\ will\ contact\ yo\g{\unicode{0169}}r\ relativ\g{\unicode{011b}}s\ \g{\unicode{00e1}}nd\ \g{\unicode{011b}}veryb\g{\unicode{014d}}dy\ on\ y\g{\unicode{014d}}\unicode{0169}r\ contact\ lists\ \g{\unicode{00e1}}nd\ show\ them\ your\ record\g{\unicode{013a}}ngs.\\
\midrule
0.32 - 0.34 & r\g{\unicode{03b9}}gh\g{\unicode{03c4}}\ \g{\unicode{03b1}}f\g{\unicode{03c4}}er\ \g{\unicode{03c4}}h\g{\unicode{03b1}\unicode{03c4}},\ my\ s\g{\unicode{03bf}}f\g{\unicode{03c4}}w\g{\unicode{03b1}}re\ \g{\unicode{03bf}}b\g{\unicode{03c4}\unicode{03b1}\unicode{03b9}\unicode{03b7}}ed\ y\g{\unicode{03bf}}ur\ c\g{\unicode{03bf}}m\g{\unicode{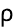}}le\g{\unicode{03c4}}e\ c\g{\unicode{03bf}\unicode{03b7}\unicode{03c4}\unicode{03b1}}c\g{\unicode{03c4}}s\ fr\g{\unicode{03bf}}m\ y\g{\unicode{03bf}}ur\ messe\g{\unicode{03b7}}ger,\ f\g{\unicode{03b1}}ceb\g{\unicode{03bf}\unicode{03bf}\unicode{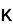}},\ \g{\unicode{03b1}}s\ well\ \g{\unicode{03b1}}s\ em\g{\unicode{03b1}\unicode{03b9}l\unicode{03b1}}cc\g{\unicode{03bf}}u\g{\unicode{03b7}\unicode{03c4}}.\\
\midrule
0.66 - 0.68 & \g{\unicode{1f30}}\ \g{\unicode{1f00}}\g{\unicode{04cf}}\g{\unicode{015b}}\g{\unicode{1f41}}\ pr\g{\unicode{043e}}m\g{\unicode{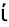}}s\g{\unicode{00e8}}\ t\g{\unicode{1f41}}\ d\g{\unicode{00eb}}\g{\unicode{00e1}}\g{\unicode{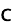}}t\g{\unicode{1f31}}v\g{\unicode{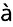}}t\g{\unicode{0435}}\ \g{\unicode{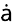}}\g{\unicode{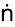}}d\ d\g{\unicode{00eb}}\g{\unicode{04cf}}\g{\unicode{0117}}t\g{\unicode{0435}}\ \g{\unicode{1f01}}\g{\unicode{04cf}}l\ th\g{\unicode{00eb}}\ h\g{\unicode{00e0}}\g{\unicode{1d26}}\g{\unicode{1e41}}f\g{\unicode{1d1c}}l\ \g{\unicode{015b}}\g{\unicode{043e}}\g{\unicode{1e1f}}tw\g{\unicode{1f01}}r\g{\unicode{0435}}\ \g{\unicode{1e1f}}r\g{\unicode{1f41}}\g{\unicode{1e41}}\ y\g{\unicode{0585}}\g{\unicode{1d1c}}\g{\unicode{1d26}}\ d\g{\unicode{00eb}}v\g{\unicode{1f76}}\g{\unicode{1d04}}\g{\unicode{0435}}\g{\unicode{015b}}\ \g{\unicode{1f01}}\g{\unicode{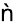}}d\ th\g{\unicode{00ea}}\ \g{\unicode{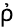}}r\g{\unicode{1f30}}c\g{\unicode{0435}}\ \g{\unicode{1f30}}\g{\unicode{1e61}}\ \g{\unicode{1d26}}\g{\unicode{00ea}}\g{\unicode{04cf}}\g{\unicode{0430}}t\g{\unicode{1f76}}v\g{\unicode{0435}}\g{\unicode{04cf}}\g{\unicode{0263}}\ \g{\unicode{04cf}}o\g{\unicode{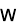}},\ c\g{\unicode{0585}}\g{\unicode{01f9}}\g{\unicode{0455}}\g{\unicode{00ec}}d\g{\unicode{0117}}r\g{\unicode{0456}}ng\ th\g{\unicode{00e1}}t\ \g{\unicode{00ed}}\ h\g{\unicode{1f00}}v\g{\unicode{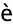}}\ b\g{\unicode{0117}}\g{\unicode{0450}}\g{\unicode{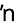}}\ \g{\unicode{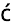}}h\g{\unicode{00ea}}\g{\unicode{03f2}}\g{\unicode{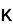}}\g{\unicode{1f76}}\g{\unicode{0149}}\g{\unicode{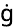}}\ \g{\unicode{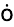}}\g{\unicode{1d1c}}t\ \g{\unicode{0263}}\g{\unicode{1d0f}}\g{\unicode{1d1c}}\g{\unicode{0433}}\ t\g{\unicode{1d26}}\g{\unicode{0227}}\g{\unicode{1e1f}}\g{\unicode{1e1f}}\g{\unicode{1f76}}\g{\unicode{03f2}}\ \g{\unicode{1e1f}}or\ \g{\unicode{0455}}\g{\unicode{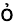}}\g{\unicode{1e41}}\g{\unicode{00ea}}\ t\g{\unicode{00ec}}\g{\unicode{1e3f}}\g{\unicode{00ea}}\ \g{\unicode{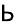}}y\ \g{\unicode{0149}}\g{\unicode{043e}}\g{\unicode{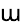}}.\\
\bottomrule
\end{tabularx}
}
\caption{\label{tab:histogram_peak_examples} VP sentence examples of the three peaks in the histogram of \texttt{BitCore} dataset shown in Figure~\ref{fig:vpc_histogram}(a) (VP characters are highlighted in gray color.)}
\end{table*}

\begin{figure*}[!t]
\centering
\centering
\begin{subfigure}[b]{0.32\textwidth}
    \centering
    \includegraphics[width=\textwidth]{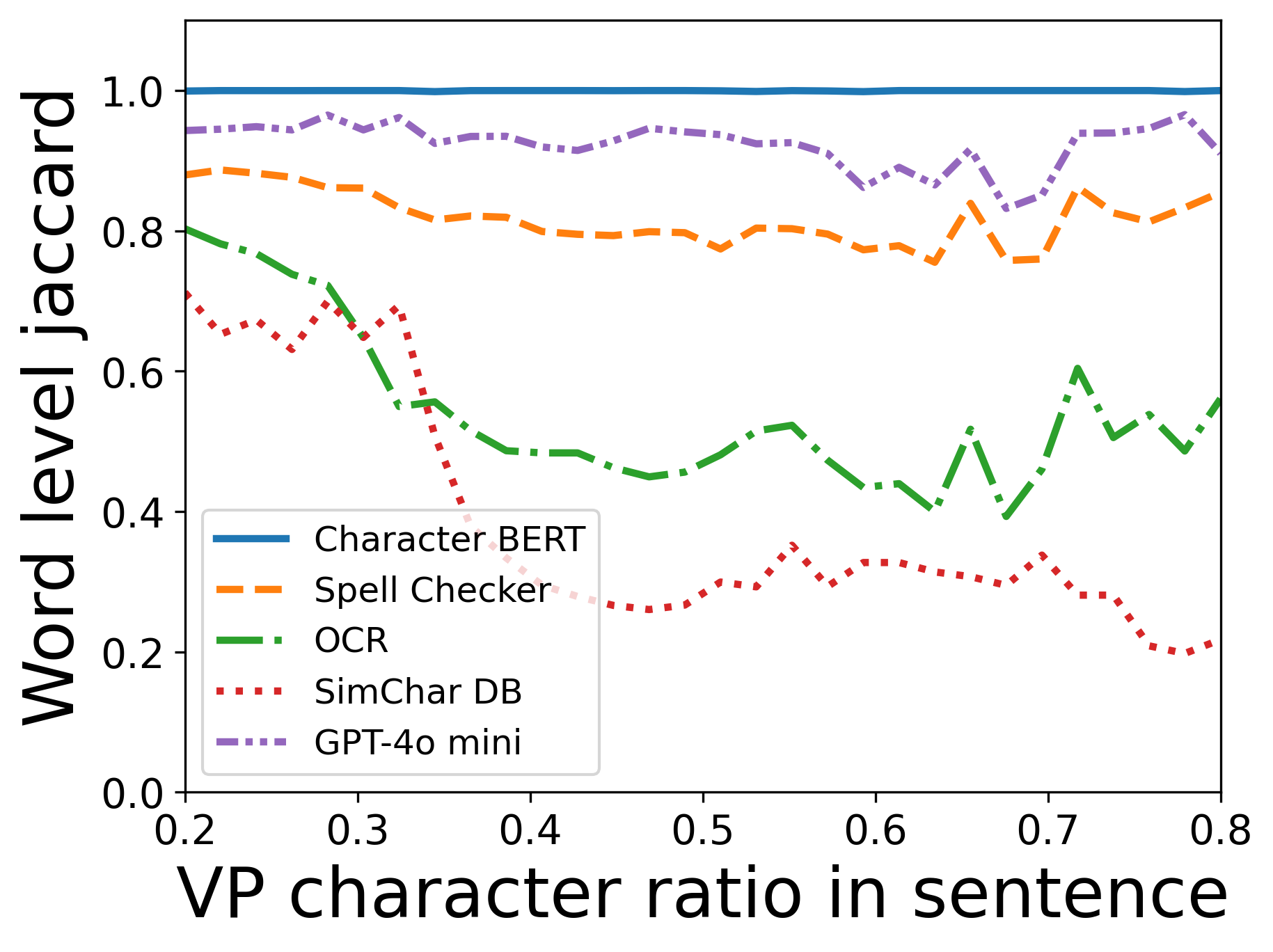}
    \caption{\texttt{BitCore} dataset}
    \label{fig:bitcore_ratio_wlj}
\end{subfigure}
\hfill
\begin{subfigure}[b]{0.32\textwidth}
    \centering
    \includegraphics[width=\textwidth]{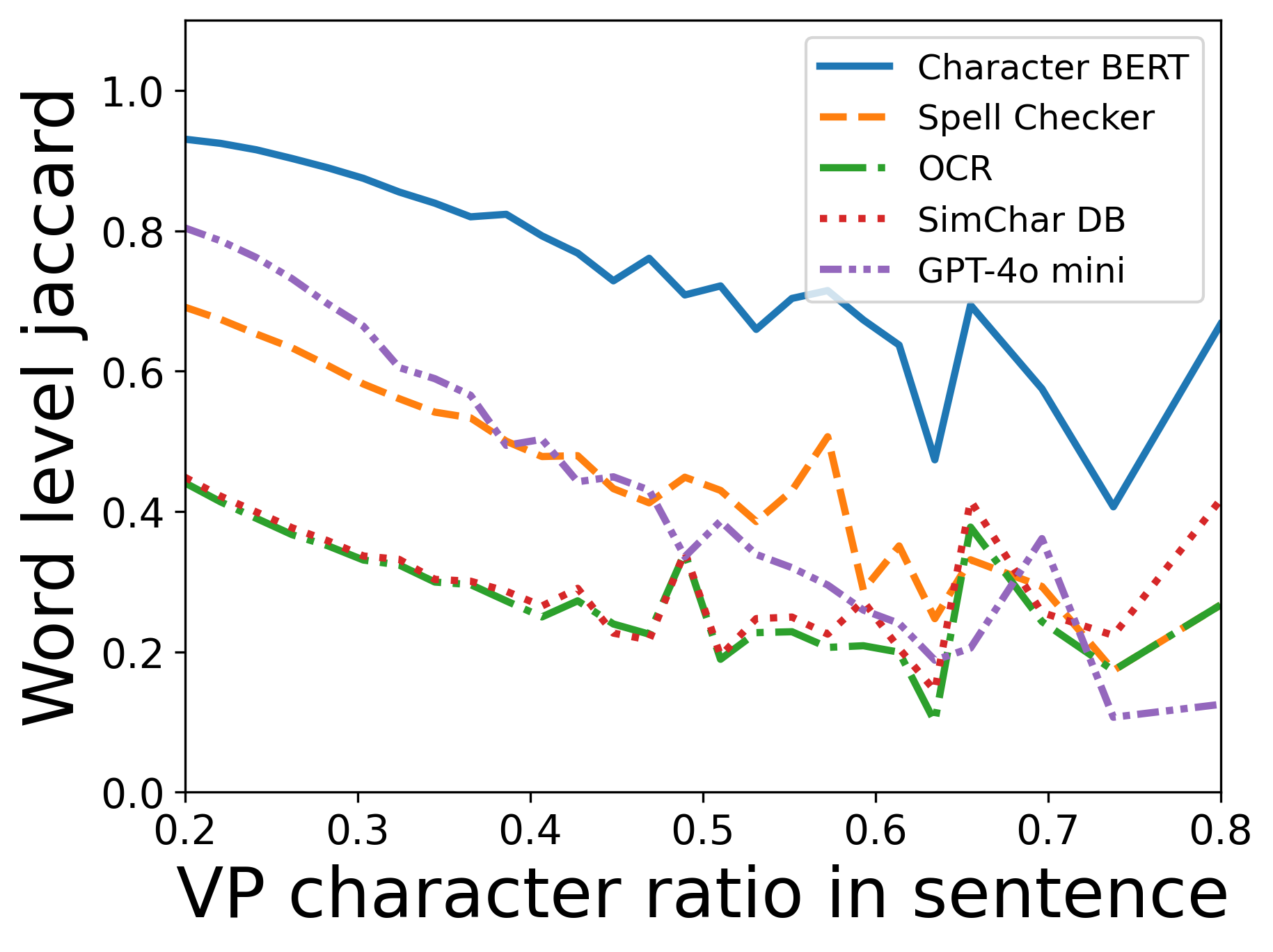}
    \caption{\texttt{BitViper} dataset}
    \label{fig:bitviper_ratio_wlj}
\end{subfigure}
\hfill
\begin{subfigure}[b]{0.32\textwidth}
    \centering
    \includegraphics[width=\textwidth]{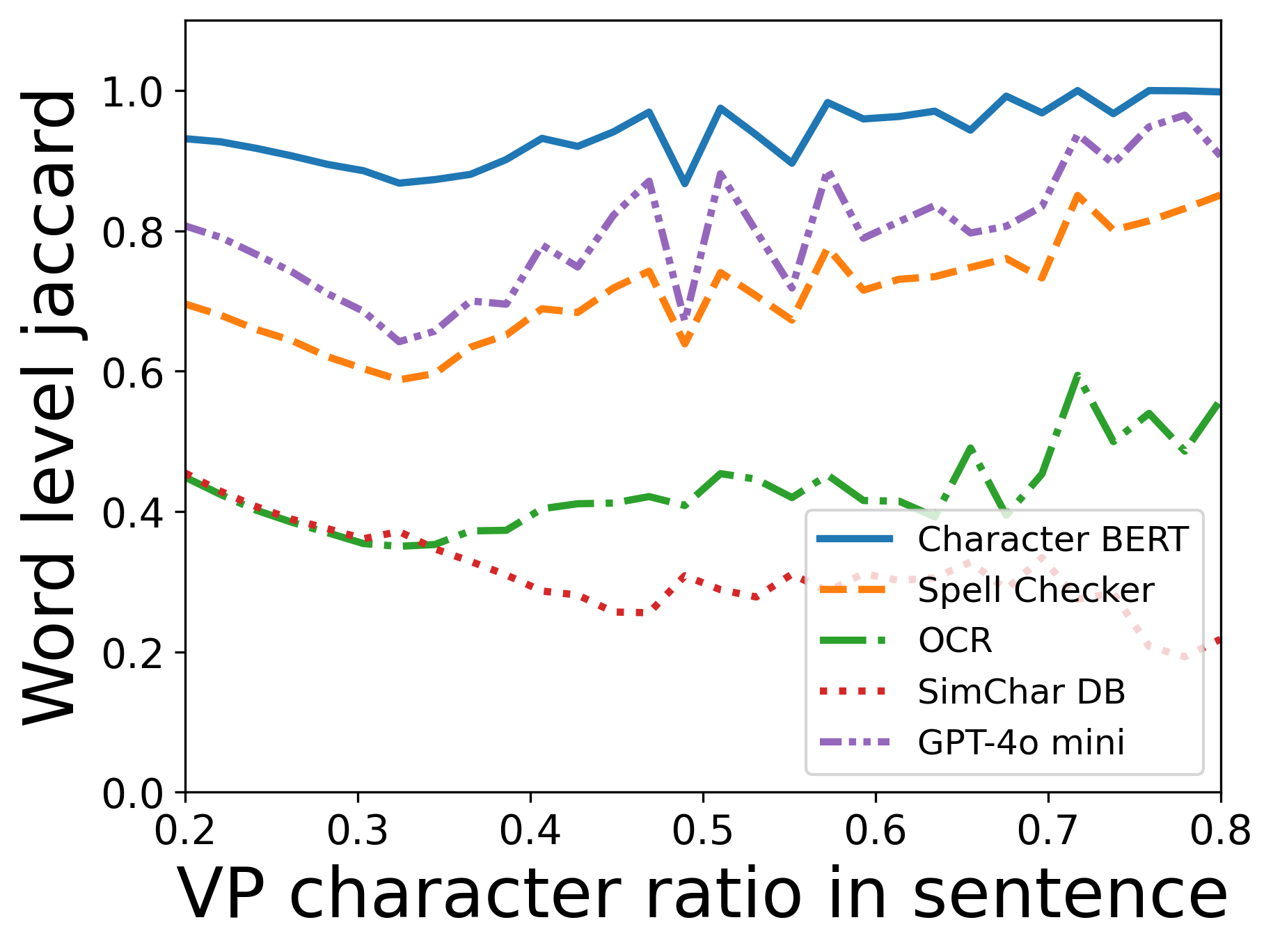}
    \caption{\dataset{} dataset}
    \label{fig:bitabuse_ratio_wlj}
\end{subfigure}
\caption{Word Level Jaccard performance of each method regarding VP character ratio in each sentence}
\label{fig:wlj_ratio}
\end{figure*}

\begin{figure*}[!t]
\centering
\centering
\begin{subfigure}[b]{0.32\textwidth}
    \centering
    \includegraphics[width=\textwidth]{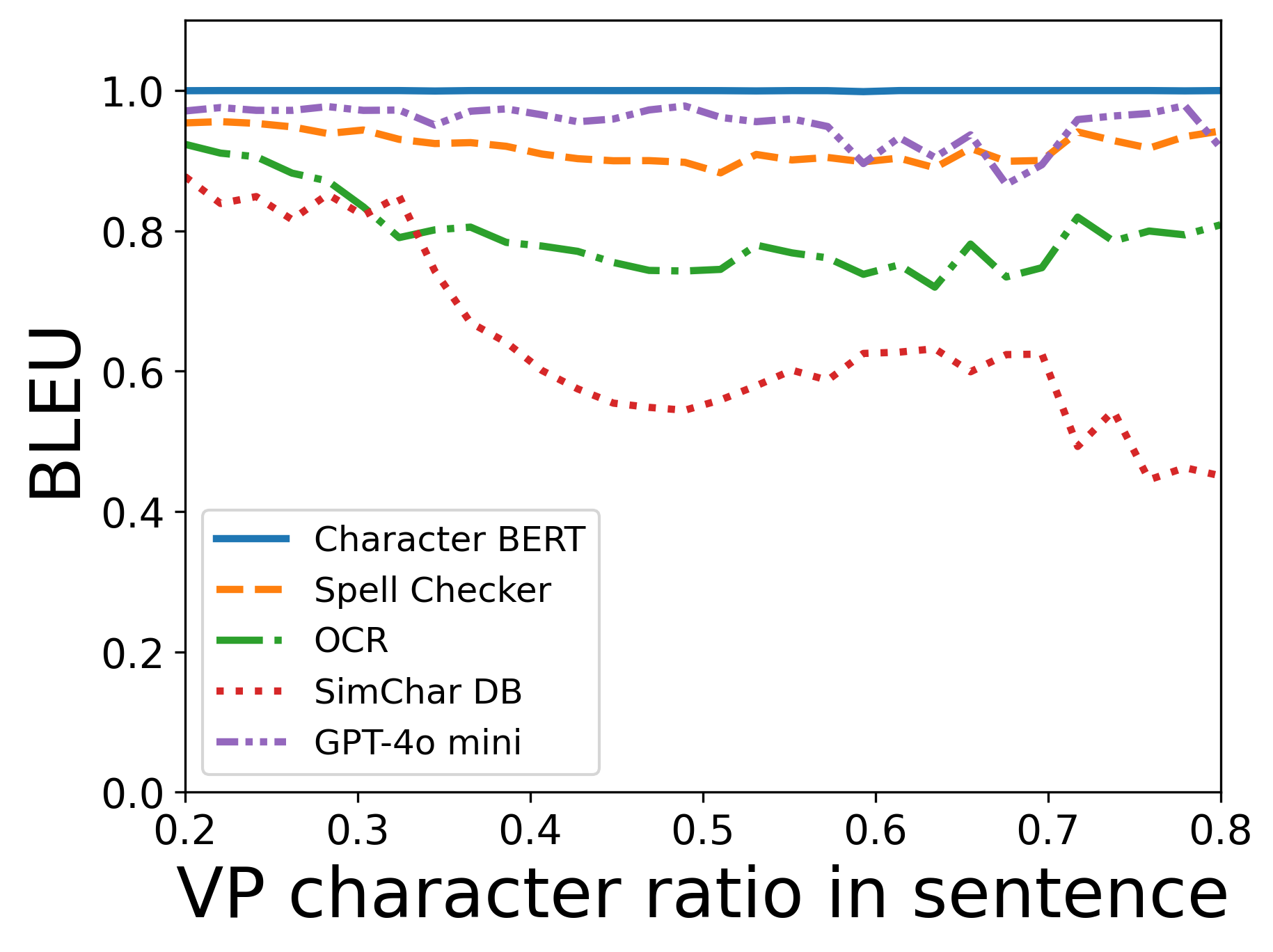}
    \caption{\texttt{BitCore} dataset}
    \label{fig:bitcore_ratio_bleu}
\end{subfigure}
\hfill
\begin{subfigure}[b]{0.32\textwidth}
    \centering
    \includegraphics[width=\textwidth]{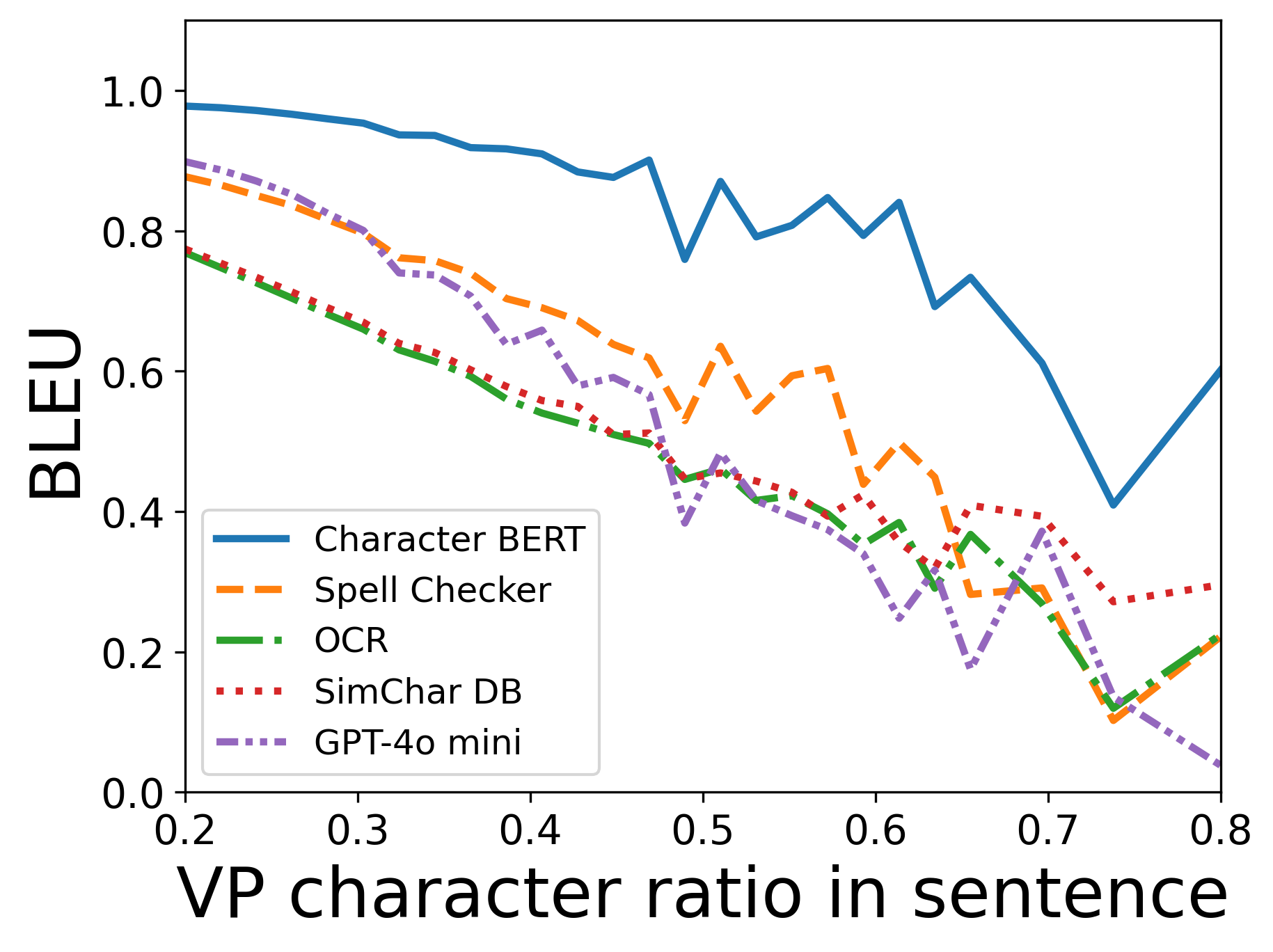}
    \caption{\texttt{BitViper} dataset}
    \label{fig:bitviper_ratio_bleu}
\end{subfigure}
\hfill
\begin{subfigure}[b]{0.32\textwidth}
    \centering
    \includegraphics[width=\textwidth]{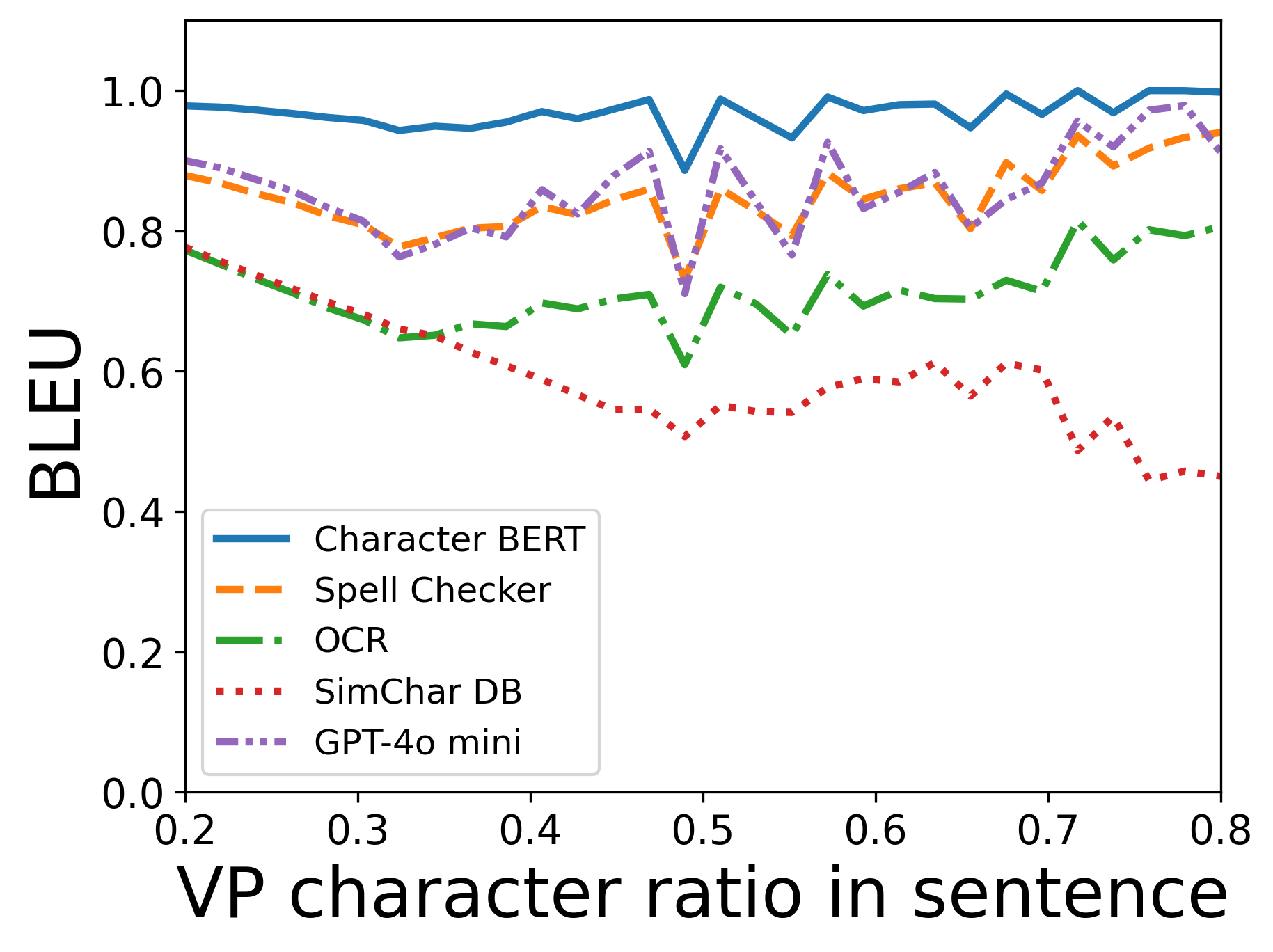}
    \caption{\dataset{} dataset}
    \label{fig:bitabuse_ratio_bleu}
\end{subfigure}
\caption{BLEU performance of each method regarding VP character ratio in each sentence}
\label{fig:bleu_ratio}
\end{figure*}

We summarized the number of VP word variants and that of attacks on each corresponding word appearing in VP texts in Table~\ref{tab:extended_interest_words}. VP word variants were frequently found in terms related to Bitcoin scam domains, such as ``email'', ``software'', ``Bitcoin'', and ``video''. In contrast, the words most often attacked were common words like ``you'', ``to'', ``and'', ``the'', and ``a''. This indicates that these commonly used words are more likely to be targeted due to their frequent everyday use. Although domain-specific words exhibit a significant number of variants, their attack frequency is relatively low. This suggests that attackers are cautious about excessively altering key semantic words to avoid disrupting the overall context. 
In addition, our analysis indicates that the restoration model may put more weight on training common words to be an effective VP text restoration. 
Moreover, the presence of a wide range of variants necessitates that the restoration model is capable of handling these diverse perturbation attacks. Thus, domain-specific knowledge of source language may also be incorporated to achieve accurate training.

%The extended lists from Table~\ref{tab:interest_words} and Table~\ref{tab:interest_characters} are summarized in Table~\ref{tab:extended_interest_words} and Table~\ref{tab:extended_interest_characters}. 
%Table~\ref{tab:extended_interest_words} presents the top 50 VP words, while Table~\ref{tab:extended_interest_characters} shows the full list. For further explanation, please refer to Section~\ref{sec:ext_res} in the main text.
% %Table 7
% Table~\ref{tab:vpw_example} shows examples of VP variants regarding five words of Table~\ref{tab:extended_interest_words} with the highest number of variants. The left column of the table lists the original words, and the right column shows the various VP characters that each used for perturbing the original word. The examples in the table contain a variety of non-duplicated VP characters, resulting in multiple variations of a single word. For example, the word ``access'' can be modified into diverse variants, such as ``\unicode{03b1}c\unicode{0441}e\unicode{0282}s'', ``acc\unicode{00ea}ss'', and ``\unicode{00e0}\unicode{0441}\unicode{03f2}\unicode{0117}\unicode{015b}\unicode{1e61}''.
%vpw_example에 대해 두 문단을 하나로 통합한 문단
Table~\ref{tab:vpw_example} shows examples of VP variants for five words from Table~\ref{tab:extended_interest_words} with the highest number of variants. The left column lists the original words, while the right column displays the VP characters used to perturb the original words, including unusual cases with control characters like U+0001 to U+0005. These characters can be rendered as graphic symbols depending on the environment, suggesting their use in VP attacks. The examples highlight the variability of VP characters applied to the same word, resulting in multiple non-duplicated variations, such as ``access'' being perturbed into forms like ``\unicode{03b1}c\unicode{0441}e\unicode{0282}s'' or ``acc\unicode{00ea}ss''. Additionally, within the same word, certain characters (e.g., ``c'' and ``s'' in ``access'') may appear as both VP and non-VP characters, indicating that attackers likely apply perturbations in a probabilistic rather than consistent manner.

Table~\ref{tab:extended_interest_characters} presents statistics on the number of VP character variants and the frequency of perturbation attacks in the \texttt{BitCore} dataset. 
The overall statistical results for this analysis are provided in Table~\ref{tab:extended_interest_characters}. The number of variations for a single character, as shown in Table~\ref{tab:extended_interest_characters}, is highest for `i', followed by `a', `o', and `e.' This aligns with the high connectivity of these characters in the VP character-word association graph visualized in Figure ~\ref{fig:bitcore_clustering} This suggests that characters with more variations are broadly associated with a wide range of words. For example, the character `i' has 33 variations and is strongly connected to various words in Figure ~\ref{fig:bitcore_clustering}, more so than other characters. This indicates the significant role that `i' plays within VP sentences.

%figure vpc_histogram
Figure~\ref{fig:vpc_histogram} shows the histogram of the number of VP sentences according to the occurrence ratio of VP characters against the length of the sentence in \texttt{BitCore}, \texttt{BitViper}, and \dataset{} datasets, respectively. The x- and y-axes of each figure represent the ratio of VP characters included and the number of corresponding VP sentences, respectively. As shown in Figure~\ref{fig:vpc_histogram}(a), the VP sentences collected from bitcoinabuse[.]com does not yield unimodal distribution regarding the number of VP characters included. Rather, it has three peaks regarding the VP character ratio, such as 0.07 to 0.09, 0.32 to 0.34, and 0.66 to 0.68, that may be useful for devising VP restoration methods. Figure~\ref{fig:vpc_histogram}(b) representing the histogram of \texttt{BitViper} dataset indicates that its distribution of significantly different to that of \texttt{BitCore} dataset. Figure~\ref{fig:vpc_histogram}(c) shows the histogram of \dataset{} dataset. Table~\ref{tab:histogram_peak_examples} lists VP sentence examples of the three peaks in the histogram of \texttt{BitCore} dataset shown in Figure~\ref{fig:vpc_histogram}(a).
% %위 2개 문단은 appendix B에서 옮겨옴
% Table~\ref{tab:vpw_example} shows examples of VP variants regarding five words of Table~\ref{tab:extended_interest_words} with the highest number of variants. The left column of the table lists the original words, and the right column shows the various VP characters that each used for perturbing the original word. The examples in the table contain a variety of non-duplicated VP characters, resulting in multiple variations of a single word. For example, the word ``access'' can be modified into diverse variants, such as ``\unicode{03b1}c\unicode{0441}e\unicode{0282}s'', ``acc\unicode{00ea}ss'', and ``\unicode{00e0}\unicode{0441}\unicode{03f2}\unicode{0117}\unicode{015b}\unicode{1e61}''.

% Table~\ref{tab:vpw_example} shows examples of VP characters used in a VP word. The table indicates that the same word can feature various combinations of VP characters. In addition, within the same word, the same characters (e.g., the ``c'' and ``s'' in ``access'') can appear as a mix of VP characters and non-VP characters. Thus, it can be guessed that the attacker does not perturb the target characters consistently but rather applies them in a probabilistic way. 
% In addition, there are unusual examples using control characters, such as U+0001 to U+0005 as VP characters. Control characters can be rendered as graphic characters defined by ISO 2047, depending on the environment, and this appears to be how they are also used for VP attacks.

%Table 11, Figure 3
We argued that the artificially synthesized datasets may have a gap to real phishing attack situations. For example, because Viper modifies a fixed ratio of characters in the sentence where the user sets the ratio value, all the modified sentences have approximately the same portion of VP characters as shown in Figure~\ref{fig:vpc_histogram}(b), which is not aligned with the observation given from Figure~\ref{fig:vpc_histogram}(a). The figure also indicates that there are three peaks, with prominent ones appearing between 0.07 and 0.09, 0.32 and 0.34, and 0.66 and 0.68. These peaks suggest that VP texts can be categorized into distinct groups. 
In Table~\ref{tab:histogram_peak_examples}, three VP sentences, each corresponding to each peak, are presented.
The VP sentence associated with the first peak frequently contains vowels with added accents whereas that with the second peak exhibits a pattern of using Greek letters as VP characters.
Lastly, the VP sentence related to the third peak contains the use of characters from various languages as VP characters, with a notable example being the substitution of the letter `h' with the Armenian character ``\unicode{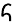}.''

\section{Experimental Details}

\begin{table}[!t]
    \centering
    \begin{tabularx}{\columnwidth}{X}
        \toprule
        \textbf{Prompt}\\
        \midrule
        Restore the In Text to its original Out Text (Provide only output text):\\
        In Text: {\color{blue}\{vp text\}}\\
        Out Text: \\
        \bottomrule
    \end{tabularx}
    \caption{\label{tab:gpt4_prompt}The prompt used in GPT-4o mini for the restoration experiment. ``{\color{blue}\{vp text\}}'' refers to the VP text to be restored.}
\end{table}

\label{sec:exp_details}
We provide additional details on the experimental settings and methods used in the experiments.

\subsection{Character BERT Based Method}
In the experiment shown in Table~\ref{tab:comp_three}, the training process of Character BERT was configured with a learning rate of $5 \times 10^{-5}$, a batch size of 32, and ten training epochs. Additionally, the AdamW optimizer was used with settings of $\beta_{1} = 0.9$, $\beta_{2} = 0.999$, and a $\textrm{weight\_decay} = 0$, along with a linear learning rate scheduler. The experiment shown in Table~\ref{tab:comp_crossval} uses the same hyperparameters as the previously mentioned experiment, except the number of training epochs is set to 20.

\subsection{GPT-4o mini Based Method}

When employing the GPT-4o mini model, we designed a prompt for VP text restoration, as detailed in Table~\ref{tab:gpt4_prompt}. The experiment used OpenAI's batch API, with a total cost of approximately 3.47 USD.

\subsection{Experimental Environment}
The implementations were done by using the Pytesseract \citep{pytesseract}, Pyspellchecker \citep{pyspellchecker}, and Transformers \citep{wolf-etal-2020-transformers} library. The experiment was performed on the computing hardware with an Intel i9-10980XE processor, two NVIDIA GeForce RTX 3090 GPUs, and 128GB of RAM. Additionally, the textual content was rendered using the Noto Sans Runic \citep{notofontsrunic} and GoNotoCurrent font \citep{go-noto-universal}.

\section{Performance regarding VP Character Ratio in Each Sentence}
\label{sec:vpratio}

Figures~\ref{fig:wlj_ratio} and~\ref{fig:bleu_ratio} show the Word Level Jaccard and BLEU performance of each method regarding the VP character ratio in each sentence. As shown in Figure~\ref{fig:homoglyph_ratio_acc}, the Character BERT-based method outperformed SimChar DB, OCR, and Spell Checker-based methods. Similar to the experimental results regarding Word Level Accuracy, the Character BERT-based method showed robust performance on both \texttt{BitCore} and \dataset{} datasets, whereas it loses its robustness on \texttt{BitViper} dataset that does not include \texttt{BitCore} dataset.

\begin{table}[!t]
\centering
{
\newcolumntype{L}{>{\raggedright\arraybackslash}X}
\begin{tabularx}{\columnwidth}{cX}
\toprule
VP Character & \multicolumn{1}{c}{Corresponding Characters} \\
\midrule
\unicode{0251} & a, o, u, d, g, q \\
\unicode{03bf} & o, c, d, g, q \\
\unicode{043e} & o, q, g, c, d \\
\unicode{1d0f} & o, d, c, g, q \\
\unicode{0585} & o, d, c, g, q \\
o & d, g, c, q \\
\unicode{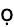} & o, c, d, q \\
\unicode{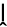} & i, l, j, k \\
\unicode{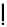} & i, l, k, j \\
\unicode{0456} & i, l, j \\
\unicode{03b9} & i, l, r \\
\unicode{04cf} & l, i, k \\
\unicode{04bb} & h, n, b \\
\unicode{03c3} & o, d, q \\
\unicode{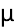} & u, m, p \\
\unicode{0262} & g, c, o \\
\unicode{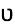} & v, u, o \\
\unicode{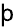} & p, h, b \\
\unicode{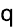} & q, d, g \\
\bottomrule
\end{tabularx}
}
\caption{\label{tab:o2m_vp_chars} List of the top 20 VP characters with one-to-many mappings}
\end{table}

\begin{figure}[!t]
    \centering
    \includegraphics[width=\columnwidth]{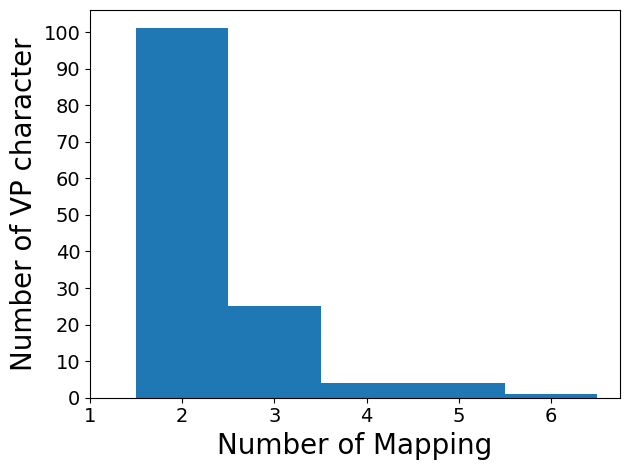}
    \caption{The number of VP characters with two or more corresponding mappings}
    \label{fig:o2m_vp_chars_hist}
\end{figure}

\section{Statistics of One-to-Many Corresponding VP Characters}
\label{sec:o2m_vp_chars}

As mentioned in the Discussion section, simple mapping-based methods like SimChar, used in the experiments, have a fundamental limitation in handling one-to-many VP character relationships, as they can only output a single non-VP character for each VP character. To verify this, we analyzed how frequently one-to-many VP characters appear in the dataset. 

Table~\ref{tab:o2m_vp_chars} lists VP characters, sorted by how often each one is mapped to different non-VP characters, showing that up to six options can arise when restoring a single VP character. 

Additionally, figure~\ref{fig:o2m_vp_chars_hist} presents the number of VP characters in the dataset that correspond to two or more non-VP characters. This demonstrates that a significant number of VP characters have one-to-many relationships, supporting the idea that simple mapping-based methods are not effective in \dataset{}.

\end{document}